\documentclass[10pt,journal,compsoc]{IEEEtran}

\pdfoutput=1\relax                   % create PDFs from pdfLaTeX
\usepackage{graphicx}                % allow us to embed graphics files
\DeclareGraphicsExtensions{.pdf,.png,.jpg,.jpeg} % for pdflatex we expect .pdf, .png, or .jpg files

\graphicspath{{figures/}{pics/}{images/}{./}} % where to search for the images
\usepackage{microtype}                 % use micro-typography (slightly more compact, better to read)
\PassOptionsToPackage{warn}{textcomp}  % to address font issues with \textrightarrow
\usepackage{textcomp}                  % use better special symbols
\usepackage{mathptmx}                  % use matching math font
\usepackage{times}                     % we use Times as the main font
\usepackage{cite}                      % needed to automatically sort the references
\usepackage{tabu}                      % only used for the table example
\usepackage{booktabs}                  % only used for the table example
\usepackage{amsmath,amssymb,amsfonts}
\usepackage{xcolor}
\usepackage{algorithm}
\usepackage{algpseudocode}
\usepackage{tablefootnote}
 \usepackage{tabularray}
\usepackage{bbding}

\newcommand{\revised}[1]{{\textcolor{black}{#1}}}

\begin{document}

\title{A Unified Particle-Based Solver for Non-Newtonian Behaviors Simulation}

\author{Chunlei Li,
	Yang Gao~\thanks{Corresponding author: gaoyangvr@buaa.edu.cn, qin@cs.stonybrook.edu.},
	Jiayi He,
	Tianwei Cheng,
	Shuai Li,
	Aimin Hao,
	Hong Qin,~\IEEEmembership{Member,~IEEE}
	\IEEEcompsocitemizethanks{%\IEEEcompsocthanksitem J. Li is with the State Key Laboratory of Virtual Reality Technology and Systems, Beihang University, Beijing 100191, China.

		\IEEEcompsocthanksitem C. Li, Y. Gao, J. He, T. Cheng, S. Li and A. Hao are with the State Key Laboratory of Virtual Reality Technology and Systems, and the Beijing Advanced Innovation Center for Biomedical Engineering, Beihang University, Beijing 100191, China, and also with the Research Unit of Virtual Body and Virtual Surgery (2019RU004), Chinese Academy of Medical Sciences, Beijing 100050, China.

		\IEEEcompsocthanksitem S. Li and A. Hao are with Peng Cheng Laboratory, Shenzhen 518066, China.
		\IEEEcompsocthanksitem S. Li is also with Zhongguancun Laboratory, Beijing 100094, China.
		\IEEEcompsocthanksitem H. Qin is with the Department of Computer Science, Stony Brook University (SUNY at Stony Brook), New York 11794-2424, USA.
USA}% <-this % stops a space

	}

\IEEEtitleabstractindextext{%
	\begin{abstract}
		In this paper, we present a unified framework to simulate non-Newtonian behaviors. We combine viscous and elasto-plastic stress into a unified particle solver to achieve various non-Newtonian behaviors ranging from fluid-like to solid-like. Our constitutive model is based on a Generalized Maxwell model, which incorporates viscosity, elasticity and plasticity in one non-linear framework by a unified way. On the one hand, taking advantage of the viscous term, we construct a series of strain-rate dependent models for classical non-Newtonian behaviors such as shear-thickening, shear-thinning, Bingham plastic, etc. On the other hand, benefiting from the elasto-plastic model, we empower our framework with the ability to simulate solid-like non-Newtonian behaviors, i.e., visco-elasticity/plasticity. In addition, we enrich our method with a heat diffusion model to make our method flexible in simulating phase change. Through sufficient experiments, we demonstrate a wide range of non-Newtonian behaviors ranging from viscous fluid to deformable objects. We believe this non-Newtonian model will enhance the realism of physically-based animation, which has great potential for computer graphics.%
	\end{abstract}

	}

\maketitle

\IEEEpeerreviewmaketitle

\IEEEdisplaynontitleabstractindextext

% ------------------------------------------------------------------
%                                   main body
% ------------------------------------------------------------------

% ------------------------------------------------------------------
%                              section Introduction                              %
% ------------------------------------------------------------------

\section{Introduction}

From melting ice cream and ketchup to human blood, non-Newtonian materials are everywhere in our daily lives. In contrast to Newtonian fluids, non-Newtonian fluids display a non-linear relationship between stress and strain rate, making them versatile and exhibiting variable properties that are different from those of traditional solids and fluids. In fact, non-Newtonian materials can be seen as materials in-between solids and fluids and therefore possess characteristics of both. By simulating these interesting non-Newtonian behaviors, we can create better physically-based animations for computer graphics applications.

Traditionally, there are two routes to simulate non-Newtonian materials in existing research. The first approach is to use a highly viscous fluid to mimic solid-like behaviour. This type of work dates back to Stora~\cite{Stora1999}. The key to success in this approach is to design a stable viscosity solver that can withstand drastic changes in viscosity~\cite{Peer2015,Goldade2019-Adaptive-Octree-Viscosity}. \revised{However, these methods are limited by viscous stress. Regardless of how large the viscous stress is, it solely influences the deformation rate and not the degree of deformation, meaning that elastic effects are not expected.
Therefore, the non-Newtonian material produced by this method may not truly be “solid-like”.} The second approach is the opposite, simulating materials from the elasticity theory and making them fluid-like. Recent works in Material Point Methods (MPM) fall under this category~\cite{Stomakhin2014,Su2021,limpmnet2023}. In comparison to the traditional Finite Element Method (FEM), MPM allows for the movement of Gaussian integral points, enabling the material to flow and giving it the ability to simulate various materials with complex, variable properties, including non-Newtonian materials. However, MPM faces a numerical adhesion problem. For low-resolution simulations with a small number of grid cells, the material may adhere together. This is because all information needs to be transferred to the grid to calculate the stress and update the deformation gradient, which limits the grid resolution and thus the resolution of the entire simulation. The material points belonging to different objects will be gathered into the same grid point and will never split again. Increasing the grid resolution will alleviate the problem, but the increase in complexity will limit the computational efficiency of the simulation.

In this research, we aim to effectively simulate non-Newtonian materials with physical properties that range from viscous fluids to deformable solids. We introduce the physics-based elastic stress to the Smoothed Particle Hydrodynamics (SPH) solver~\cite{koschier2022survey}, thus combining the fluid-like resistance and the solid-like resistance within a unified framework. A notable merit of this practice compared to current viscous SPH methods is that our model is fully resistant to any non-Newtonian deformation. Moreover, we empower our method with the capability of simulating various strain rate-dependent materials, including shear-thinning, shear-thickening, Bingham plastic, and more. By utilizing our unified framework, which combines an elasto-plastic solver with a viscosity solver, we are able to achieve sufficient and realistic non-Newtonian phenomena.

\begin{figure}[htbp]
	\centering
	\includegraphics[width=0.95\linewidth]{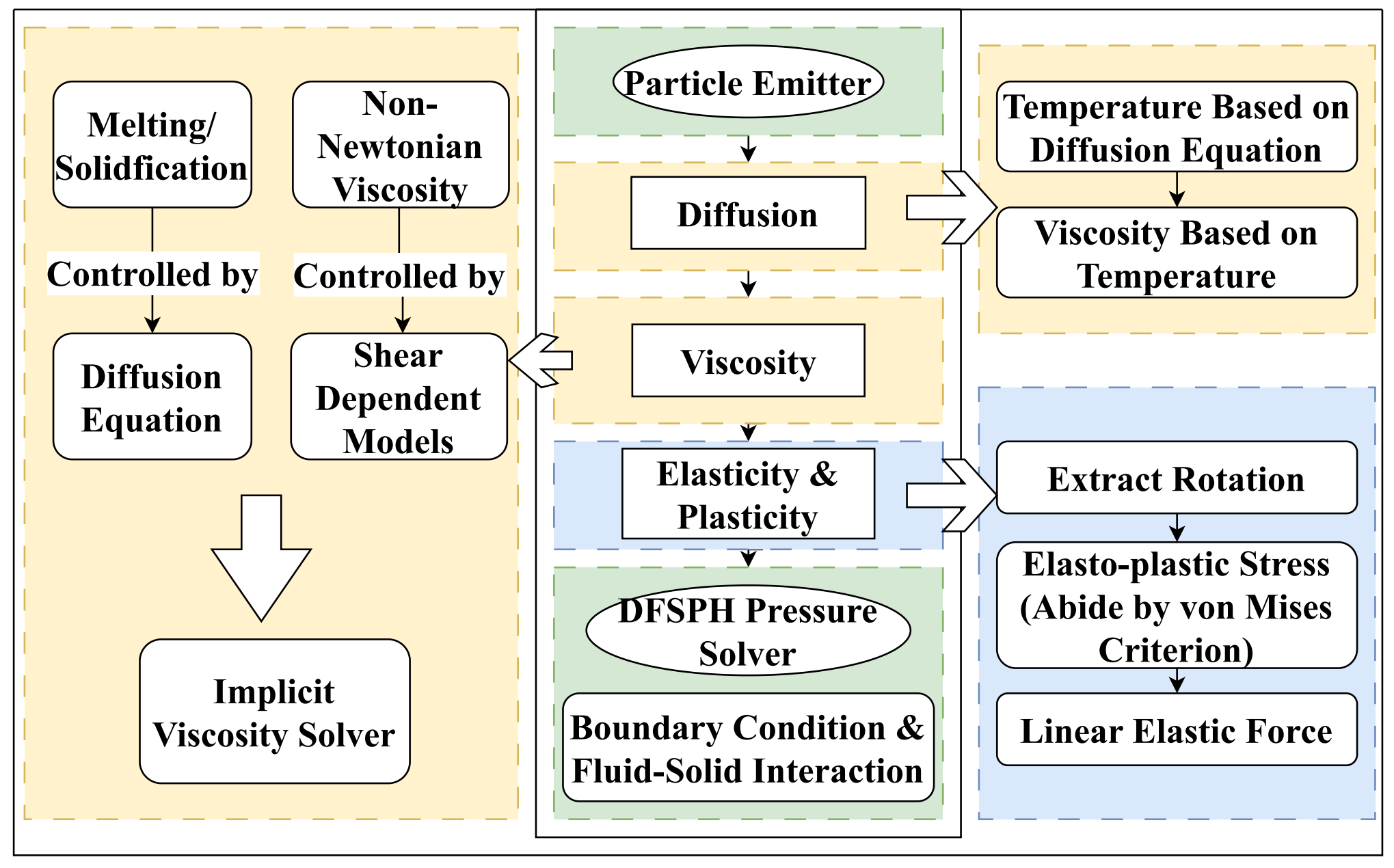}
	\caption{\revised{The pipeline of our unified framework, where the yellow background represents related to viscous model, blue background represents related to elasto-plastic model, and green background represents others.}}
	\label{fig:pipeline}
\end{figure}

Meanwhile, we also implement the diffusion equation to simulate the heat-based phase change. The pipeline is shown in Fig.~\ref{fig:pipeline}. To summarise, our salient contributions are:

\begin{itemize}
	\item We put the elastic stress and viscous stress together into the Navier-Stokes equation to develop a unified particle-based solver for simulating from fluid-like to solid-like non-Newtonian behaviors.
	\item We implement strain rate dependent non-Newtonian models, and thus many classical non-Newtonian models can be reproduced in a flexible and physically meaningful way.
	\item We combine the diffusion model as well as the visco-elastic and visco-plastic coupled models into our framework using a Generalized Maxwell Model, thus further enriching the reality and the variety of non-Newtonian phenomena.
\end{itemize}

% ---------------------------------------------------------------------------- %
%                             section related works                            %
% ---------------------------------------------------------------------------- %

\section{Related Works}

\subsection{Variable Viscosity for Non-Newtonian Material}
One of the key characteristics of non-Newtonian material is that it has variable viscosity. Variable viscosity is a long-discussed issue in computer graphics. Stora et al.~\cite{Stora1999} pioneered in this field by simulating the lava with a high and variable viscosity in 1999. \revised{Carlson et al.~\cite{Carlson2002} enhanced the numerical stability of algorithms under high viscosity for simulating phenomena, such as wax melting.} Batty et al.~\cite{Batty2008-accurate-viscous} developed a variational principle-based Eulerian fluid solver that focuses on the accuracy of complex boundary conditions of viscous flow. Larionov et al.~\cite{Larionov2017-variationalStokes} designed a Stokes flow solver that is unconditionally stable when simulating highly viscous phenomena like buckling and coiling. \revised{Goldade et al.~\cite{Goldade2019-Adaptive-Octree-Viscosity} designed an adaptive variational finite difference framework for octrees that is capable of simulating variable viscosity.}  Shao et al.~\cite{Shao-Huang2022-unsmoothed} developed a high-resolution algebraic multigrid viscosity solver capable of simulating fluids with variable viscosity.

\revised{The above-mentioned research works are based on the Eulerian viewpoint or hybrid viewpoint. Additionally, there are several studies that utilize the Lagrangian solver~\cite{Muller2004-elastic-plastic-melting,Solenthaler2007}.} The SPH method can be traced back to Monaghan~\cite{Monaghan1992}. Muller et al.~\cite{Muller03-SPH} introduced an equation-of-state based approach to enforce the incompressibility. \revised{Zhu et al.~\cite{Zhu2015-nonNewton} put forward a codimensional non-Newtonian viscosity fluid simulator based on Fluid Implicit Particle (FLIP), where Carreu-Yasuda model is used to calculate shear thinning and shear thickening.} Muller's model was further developed into a Weakly Compressible SPH (WCSPH) by Becker et al.~\cite{Becker2007-WCSPH}. Subsequently, numerous variants of SPH have been introduced to address incompressibility with greater efficiency, including the Predictive–Corrective Incompressible SPH (PCISPH)~\cite{Solenthaler2009-PCISPH}, Implicit Incompressible SPH (IISPH)~\cite{Ihmsen14-IISPH}, and Divergence-Free SPH (DFSPH)~\cite{Bender2017-DFSPH}. As for the viscosity in SPH, Koschier et al.~\cite{Koschier2019-Tut} reviewed the recent works among the SPH viscous solvers. The XSPH~\cite {Monaghan1992,Bridson2012-XSPH} is an explicit solver that takes advantage of the fact that viscosity is caused by the difference in velocities of fluid blobs. Takahashi et al.~\cite{Takahashi2015} used a second-ring neighbourhood to tackle the inconsistency issue of second-order derivatives of SPH. Peer et al.~\cite{Peer2015} decomposed the velocity gradient and reassembled it with a limited strain rate tensor to mimic the viscosity behavior. \revised{ Similarly, Peer and Teschner~\cite{Peer2017-Prescribed} used the prescribed velocity gradient to simulate the viscous fluid.}  Bender and Koschier~\cite{Bender2017-DFSPH} constrained the strain rate by a user-defined parameter to simulate viscosity. And Weiler et al.~\cite{Weiler2018-viscosity} solved the Laplacian formula in an implicit scheme, taking into account the boundary particles and treating them with their conjugate gradient solver.   \revised{Andrade et al.~\cite{Andrade2015} utilized a SPH based method to simulate the non-Newtonian viscosity with Cross model and accelerated the performance with CUDA.}

\subsection{Visco-elasticity and Visco-plasticity}
Another key characteristic of non-Newtonian material is the visco-elasticity and visco-plasticity. These materials possess properties of both the solid and liquid states~\cite{gao2021simulating}. The study of visco-elastic and visco-plastic materials has long been a topic of interest among researchers. \revised{The interest in these materials can be traced back to Terzopoulos et al. in 1989~\cite{Terzopoulos1989}, when Terzopoulos simulated visco-elastic material by manipulating the integration point of FEM.} Losasso et al.~\cite{Losasso2006} tracked the level set of different materials, including solid, so their solver is capable of simulating multiple materials with distinct and variable physical properties. Similarly, Fujisawa et al.~\cite{Fujisawa2007} adopted the Volume of Fluid method to track the boundary when simulating the ice melting. The concepts of viscosity and elasticity can be unified in terms of stress. Observing from the outcome,they both produce a limited degree of stress and hence inhibit mobility. Viscosity takes strain rate as input, whereas elasticity takes strain as input. Goktekin et al al.~\cite{Goktekin2004} took this thought further by incorporating the elastic stress into the Navier-Stokes equations in order to simulate viscoelasticity. They used this method to simulate the jelly in the Marker And Cell (MAC) framework. Muller et al.~\cite{Muller2004-elastic-plastic-melting} tackled the unified particle system to simulate elastic, plastic, and melting objects. They calculated the stress based on the traditional FEM but discretized it on particles. Becker et al.~\cite{Becker2009} followed this idea and implemented elasticity in a SPH framework.

Meanwhile, \revised{plasticity and elasticity are inseparable parts of Newtonian behaviors.} O'Brien et al.~\cite {OBrien2002} provided a solution to simulate the effect of plasticity. When the elasticity strain (or stress) exceeds a criterion, it will be constrained by the plastic law. Such a criterion is von Mises criterion in their work, and the limit is the elastic limit. A further increasing elastic strain will encounter another limit, the plastic limit. The object will fracture if the plastic limit is exceeded. A significant distinction between elasticity and plasticity is that plasticity is a history-dependent quantity that requires updating and accumulation at each time step. Muller et al.~\cite{Muller2004-elastic-plastic-melting} and Solenthaler et al. ~\cite{Solenthaler2007} adopted this method to simulate plasticity in the particle method. Another common practise to treat plasticity is to separate the deformation gradient into two components, the elastic component and the plastic component. This practise is adopted by Stomakhin et al.~\cite{Stomakhin2013-snow} during the simulation of snow, where the snow shows both elasticity and plasticity. Similar to O'Brien's work, there are two thresholds for elasticity and plasticity. \revised{Yue et al.\cite{Yue2015} used MPM to simulate the foam, which is a typical Bingham type non-Newtonian material. Their constitutive correlation is based on the Herschel-Bulkley model.} More related works of MPM ~\cite{Su2021,gao2019efficient,Fang2019-sillyRubber,Gao2018-GPU-MPM} can be classified into this type. \revised{Gissler et al.~\cite{Gissler2020-snow} simulated the snow with SPH. Their work refers to the visco-elastic model developed by Stomakhin et al.~\cite{Stomakhin2014}} \revised{Bargteil et al.~\cite{Bargteil-2007-AFE} used FEM to simulate the visco-plastic flow. To tackle the potential problem of the ill-conditioned mesh after large deformation, they used a re-mesh procedure.} \revised{Wojtan and Turk~\cite{Wojtan2008} took a similar strategy with Bargteil et al.~\cite{Bargteil-2007-AFE} to simulate the visco-elastic behavior. They carefully distinguished the situation for remeshing. With the spatial adaptivity technique, their algorithm achieves an order-of-magnitude speedup.} \revised{Ozgen et al.~\cite{Ozgen2019} simulated the behavior of shear thickening fluids with SPH model. The key factor in their method is to use a spring to capture the nonlinearity of the non-Newtonian behavior.}

\revised{For computer graphics applications, the key to depicting the visual effect of non-Newtonian behaviors is to demonstrate the material’s dynamic characteristics, which can range from fluid-like to solid-like. Therefore, a simulation algorithm must take into account the following key features: 1) variable viscosity (incorporating classical shear-related models), 2) the inclusion of elasticity, and 3) the integration of plasticity.}

\revised{
Existing research seldom covers all of the above-mentioned research issues. Early works, such as Carlson et al.~\cite{Carlson2002} and Stora et al.~\cite{Stora1999}, considered classical shear-related variable viscosity models. The work of Zhu et al.~\cite{Zhu2015-nonNewton} is among the few papers dedicated to simulating non-Newtonian behaviors, taking into account variable viscosity and elasticity. However, it only offers one shear-related variable viscosity model and lacks plasticity simulation. In recent years, some works using the MPM, such as Su et al.~\cite{Su2021} and Fang et al.~\cite{Fang2019-sillyRubber}, claim to be able to simulate non-Newtonian behaviors. However, their methods primarily focus on viscoelastic behavior. Solenthaler et al.~\cite{Solenthaler2007} proposed a unified SPH-based approach that could simulate both fluids and solids, but it did not specifically address viscoelastic materials that lie between fluids and solids. Goktekin et al.~\cite{Goktekin2004} simulated semi-fluid, semi-solid materials by adding elastic stress in the momentum equation, which offered us some inspiration. 
 However, we haven't seen specialized research that encompasses the majority of non-Newtonian behaviors, despite the fact that non-Newtonian behaviors are widely observed in real-world phenomena. As such, we believe that consolidating the three aforementioned behavioral characteristics (viscous, elastic, and plastic) into a unified framework will represent a significant contribution to the computer graphics community, at least at the application level. }

\revised{Thus we design a flexible and versatile solver to incorporate all characteristics that simulating non-Newtonian behaviors required into one unified framework.} With adjustable parameter tweaking, it is hopeful to develop more diverse strain rate-driven, strain-driven, and temperature-driven phenomena than with prior studies.

% ---------------------------------------------------------------------------- %
%                                  section GMM                                 %
% ---------------------------------------------------------------------------- %

\section{Governing equations and Generalized Maxwell Model}\label{sec:govern-eq-and-maxwell}

We present our momentum balance equation by \revised{Eq.~\ref{eq:momentum-balance}}. \revised{The primary advancement lies in the modification of the original momentum equation through the inclusion of the elasto-plastic stress $\sigma_{ep}$ within the equation, drawing inspiration from the work of Goktekin et al.\cite{Goktekin2004}.}

\begin{equation}
	\frac{D \mathbf{v}}{D t}=-\frac{1}{\rho} \nabla p+ \nabla \cdot (\sigma_v+\sigma_{ep}) +\frac{\mathbf{f_{ext}}}{\rho}, \label{eq:momentum-balance}
\end{equation}
where  $\sigma_v$ represents viscous stress, $p$ is the pressure, $\rho$ is the density, $\mathbf{f_{ext}}$ is the external force, $\sigma_{ep}$ is the elasto-plastic stress.

\revised{There are various methods to discretize and solve Eq.~\ref{eq:momentum-balance}.} In this work, we adopt the DFSPH~\cite{Bender2017-DFSPH}  for our particle-based solver, which is the most advanced SPH solver to this date~\cite{Koschier2019-Tut}.

\begin{figure}[htbp]
	\centering
	\includegraphics[width=0.4\textwidth]{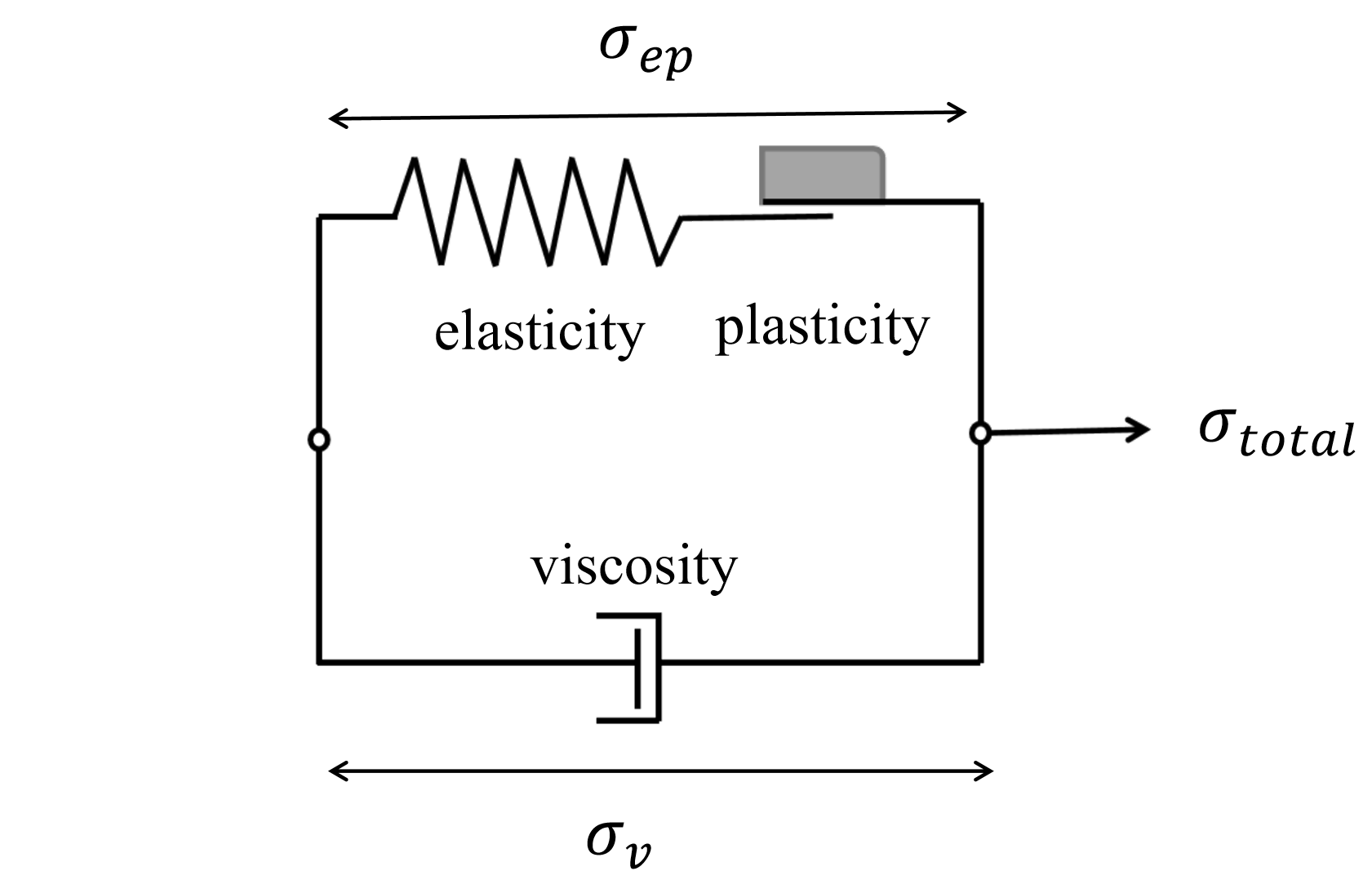}
	\caption{Generalized Maxwell model to describe one-dimensional stress-strain (strain rate) relation.}     \label{fig:maxwell}
\end{figure}

To solve Eq.~\ref{eq:momentum-balance}, we need to focus on the computation of $\sigma_v$, $\sigma_{ep}$ and their relation.

We present a Generalized Maxwell Model (GMM) to describe the constitutive relation underlying our unified non-Newtonian framework. \revised{Our inspiration is from the similar model from Fang et al.~\cite{Fang2019-sillyRubber} for viscoelasticity simulation. The Generalized Maxwell Model can be conceptualized as a combination of a spring, a slider, and a dashpot, arranged both in series and parallel. This model comprises two branches: the upper branch features a spring and a slider in series, while the lower branch is composed solely of a dashpot. Each component of the model symbolizes a distinct physical property: the spring represents elasticity, the slider embodies plasticity, and the dashpot signifies viscosity.}

\revised{Assuming that} a material undergoes a specific deformation, it will inherently exhibit some resistance to this deformation. This total resistance is represented by the total stress, denoted as $\sigma_{total}$. In our framework, this total stress can be abstracted to the sum of the resistances provided by the upper and lower branches. The resistance from the upper branch is derived from elasto-plasticity, while the resistance from the lower branch arises due to viscosity.

\begin{equation}
	\sigma_{total} = \sigma_{ep} + \sigma_v,
\end{equation}
where $\sigma_{ep}$ denotes the elasto-plastic stress, representing the resistance from the upper branch of Fig.\ref{fig:maxwell}. Similarly, $\sigma_v$ signifies the viscous stress, corresponding to the resistance provided by the lower branch in Fig.~\ref{fig:maxwell}.
\revised{Subsequently, it is necessary to determine the individual computation methods for $\sigma_{ep}$ and $\sigma_v$.}

The stress, $\sigma_{ep}$, is related to the elasto-plastic strain $\varepsilon_{ep}$, but this relation is not simply linear. The elasto-plastic strain can be further decomposed into the elastic part and the plastic part:

\begin{equation}
	\varepsilon_{ep} = \varepsilon_{e} + \varepsilon_{p}. \label{eq:addition_law_strain}
\end{equation}

\begin{figure}[htbp]
	\centering
	\includegraphics[width=0.7\linewidth]{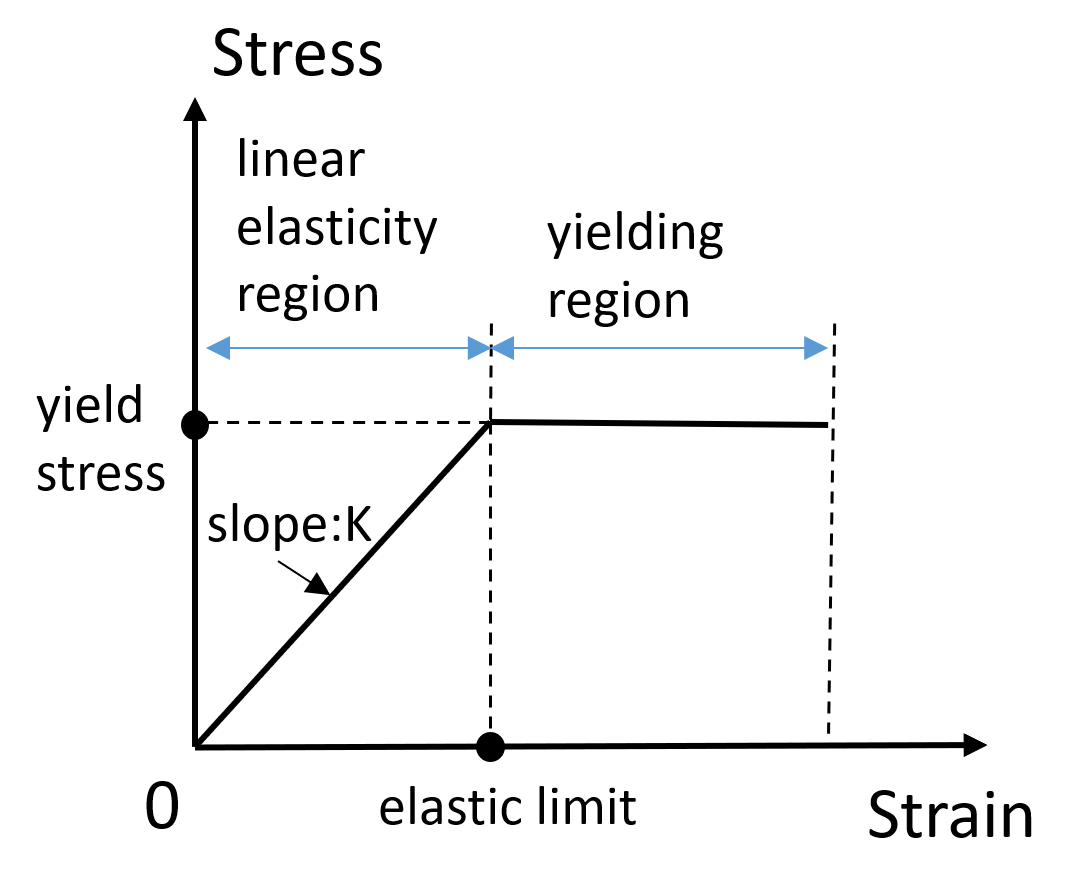}
	\caption{Schematic diagram of elasto-plastic constitutive relation in one dimension.}     \label{fig:elasto-plastic-constitution}
\end{figure}

The relation between stresses and strains for elasto-plasticity is described by Eq.\ref{eq:elasto-plastic-constituitive}.

\begin{equation}
	\begin{cases}
		\sigma_e = K:\varepsilon_e \\
		\sigma_p = 0
	\end{cases},\label{eq:elasto-plastic-constituitive}
\end{equation}
where the plastic strain $\varepsilon_{p}$ exhibits zero resistance to deformation, which results in $\sigma_{p}$ always being zero. This is a characteristic of plasticity, where the material enters a state known as yielding. In this state, the material offers no resistance or only mild resistance to deformation. This is the underlying rationale for using a slider to represent plasticity in our Generalized Maxwell model, the slider can move freely without any damping. On the other hand, the elastic constitutive relation is linear due to our adoption of linear elasticity theory, thus we use a spring to represent elasticity. In the upper row of Eq.~\ref{eq:elasto-plastic-constituitive}, $\mathit{K}$ represents a fourth-order stiffness tensor. A visual schematic diagram illustrating the stress-strain relation in one-dimensional case is presented in Fig.\ref{fig:elasto-plastic-constitution}. Please note that in this one-dimensional diagram, \textit{K} is represented as a scalar coefficient, indicating that \textit{K} represents the slope of the curve before reaching the elastic limit. Further details about elasto-plasticity are discussed in Section \ref{sec:Elasticity-Plasticity}.

The bottom branch of Fig.~\ref{fig:maxwell}, i.e., the viscosity, is symbolized by a dashpot. This is because while viscosity has a resistance effect, it differs from the elastic component in that it responds to the rate at which the deformation is applied rather than the magnitude of the deformation. This concept is encapsulated in mathematical expressions, indicating that the magnitude of the viscous stress is related to the strain rate, the first derivative of the strain with respect to time. Here, the viscous stress, $\sigma_{v}$, is determined by the viscous strain rate, $\dot{\varepsilon_{v}}$, but it's important to note that this relation is not necessarily linear.
More details of the connection between viscous stress and viscous strain rate will be discussed in Section \ref{sec:variable-viscosity}.

% ---------------------------------------------------------------------------- %
%                               section NonNewton                              %
% ---------------------------------------------------------------------------- %

\section{Non-Newtonian Materials Modeling}\label{sec:Non-Newtonian model}

Non-Newtonian behaviors can generally be classified into two categories~\cite{Phan2017,Chhabra2010}: 1) time-independent and 2) time-dependent. Our work falls under the time-dependent category due to the co-existence of viscosity and elasto-plasticity. However, we can simplify this by considering these components separately, as we have already discussed in section \ref{sec:govern-eq-and-maxwell}.
After the separation, the viscosity model can be viewed as time-independent, and we can implement it by adopting a variety of classic non-Newtonian models based on the apparent viscosity coefficient. Section \ref{sec:variable-viscosity} will discuss viscosity, and section \ref{sec:Elasticity-Plasticity} will cover elasto-plasticity. We will explain in detail how to implement these elements in our solver.

\subsection{Variable Viscosity}\label{sec:variable-viscosity}

As the viscosity may undergo drastic changes, a stable implicit solver becomes necessary. Therefore, we adopt the implicit viscosity solver from Weiler~\cite{Weiler2018-viscosity}, in which the viscous term can be discretized as follows:

\begin{equation}
	\nabla \cdot \sigma_v= 2(d+2) \nu \sum_j \frac{m_j}{\rho_j} \frac{\mathbf{v}_{i j} \cdot \mathbf{x}_{i j}}{\left\|\mathbf{x}_{i j}\right\|^2+0.01 h^2} \nabla W_{i j}\label{eq:viscous-force},
\end{equation}
where $\nu=\mu / \rho$ is the kinematic viscosity, and $\mu$ is the dynamic viscosity. And $d$ represents the dimension. $x_{ij}$ is the distance between particle $i$ and its neighbor $j$. $h$ is the support radius, $W$ is the kernel function of SPH solver. Then we impose an implicit time integration scheme for Eq.~\ref{eq:viscous-force}:

\begin{equation}
	\mathbf{v}(t+\Delta t)=\mathbf{v}^*+\frac{\Delta t}{\rho}  \nabla\cdot(\mu \nabla \mathbf{v}(t+\Delta t)).
\end{equation}

The implicit viscosity solver necessitates a distinct matrix solver. In our simulation, we employ a matrix-free conjugate gradient solver with Jacobi preconditioning. Detailed information can be found in Weiler's work~\cite{Weiler2018-viscosity}.

Thanks to the separation of viscosity and elasto-plasticity as discussed in Generalized Maxwell Model in section \ref{sec:govern-eq-and-maxwell}, we could view the viscosity as time-independent. This means the strain rate and stress can be described by a single-value function between strain rate $\dot\epsilon_v$ and viscous stress $\sigma_v$:

\begin{equation}
	\sigma_v=f\left(\dot{\varepsilon_v}\right),
\end{equation}
where $\sigma_v$ is the viscous stress, and $\dot{\varepsilon_v}$ is the viscous strain rate. This relation is non-linear so we use the $f$ to represent their relation.
Since $\mu=d\sigma_v /d\dot{\epsilon_v}$, the effective viscosity can also be expressed as the function of the strain rate, i.e. ,

\begin{equation}
	\mu=g\left(\dot{\epsilon_v}\right). \label{eq:mu_epsilon}
\end{equation}

In the following paragraphs, we will introduce six classic strain rate dependent viscosity models. Each of them can be considered a variant of Eq.~\ref{eq:mu_epsilon}. Note that both the strain rate and viscous stress are second-order tensors. However, for simplicity, we retain the convention that every time we take the power of the tensor, we first compute the Frobenius norm of that tensor.

Based on the response of stress with respect to strain rate, the time-independent non-Newtonian behaviors into three types: 1) shear-thinning body (also known as pseudo-plastic body); 2) Bingham body without shear-thinning or with shear-thinning (also known as Casson body); and 3) shear-thickening body (also known as dilatant body).

\begin{figure}[htbp]
	\centering
	\includegraphics[width=0.8\linewidth]{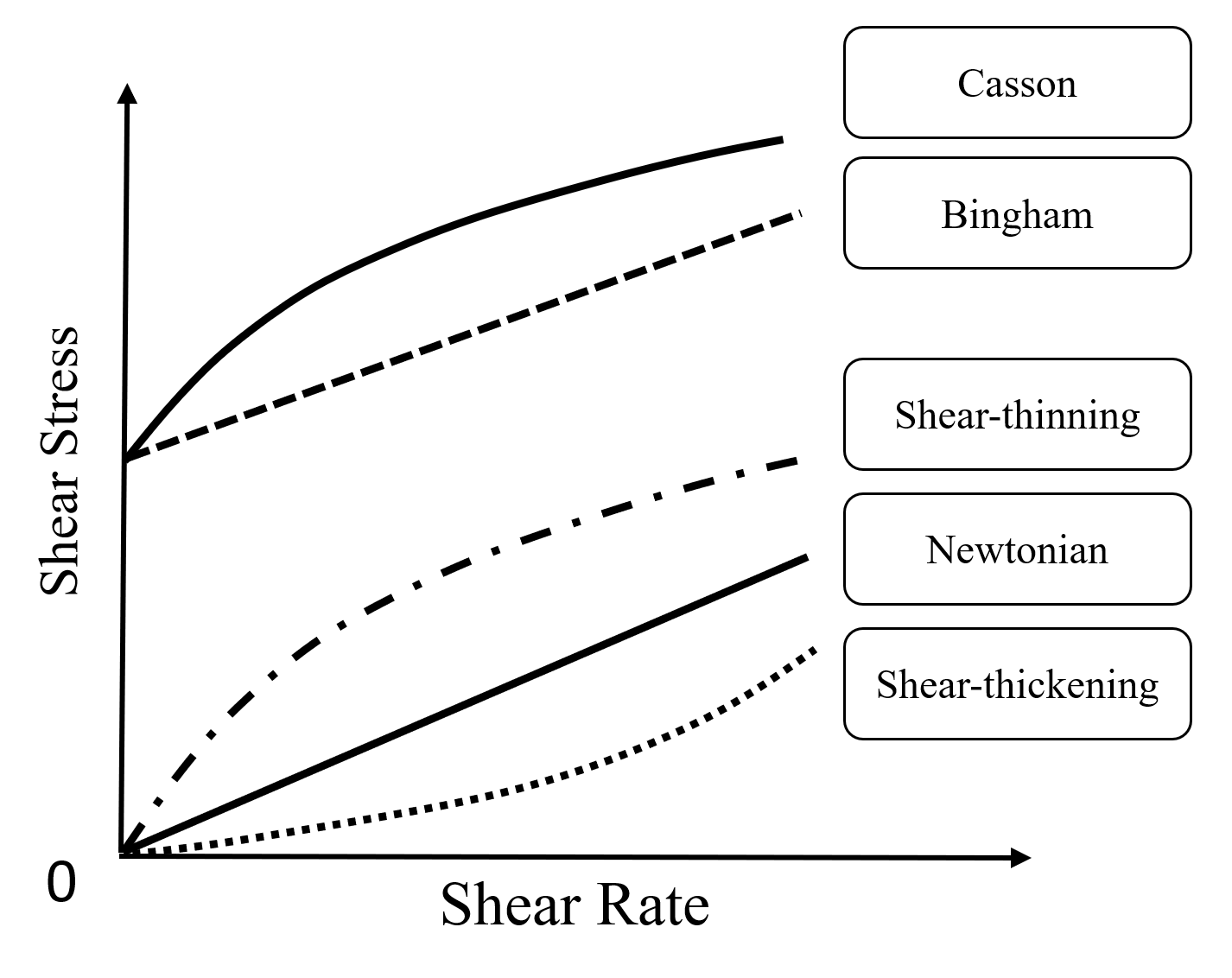}
	\caption{Classification of time-independent non-Newtonian behaviors.}
	\label{fig:rheogram}
\end{figure}

Fig.~\ref{fig:rheogram} illustrates the relation between strain rate and viscous stress for different time-independent non-Newtonian behaviors. The slope of the rheogram represents the viscosity. As it is shown, shear-thinning behavior exhibits a decreasing slope as the strain rate increases, while the opposite is observed for shear-thickening behavior. A rheogram curve of the Bingham body has a non-zero intercept on the stress axis, which is the yield stress. Shear-thinning is the most common type of time-independent, non-Newtonian behaviors. Daily necessities, such as ketchup, nail polish, and wall paint, exhibit shear-thinning behaviors.

\begin{figure}[htbp]
	\centering
	\includegraphics[width=.9\linewidth]{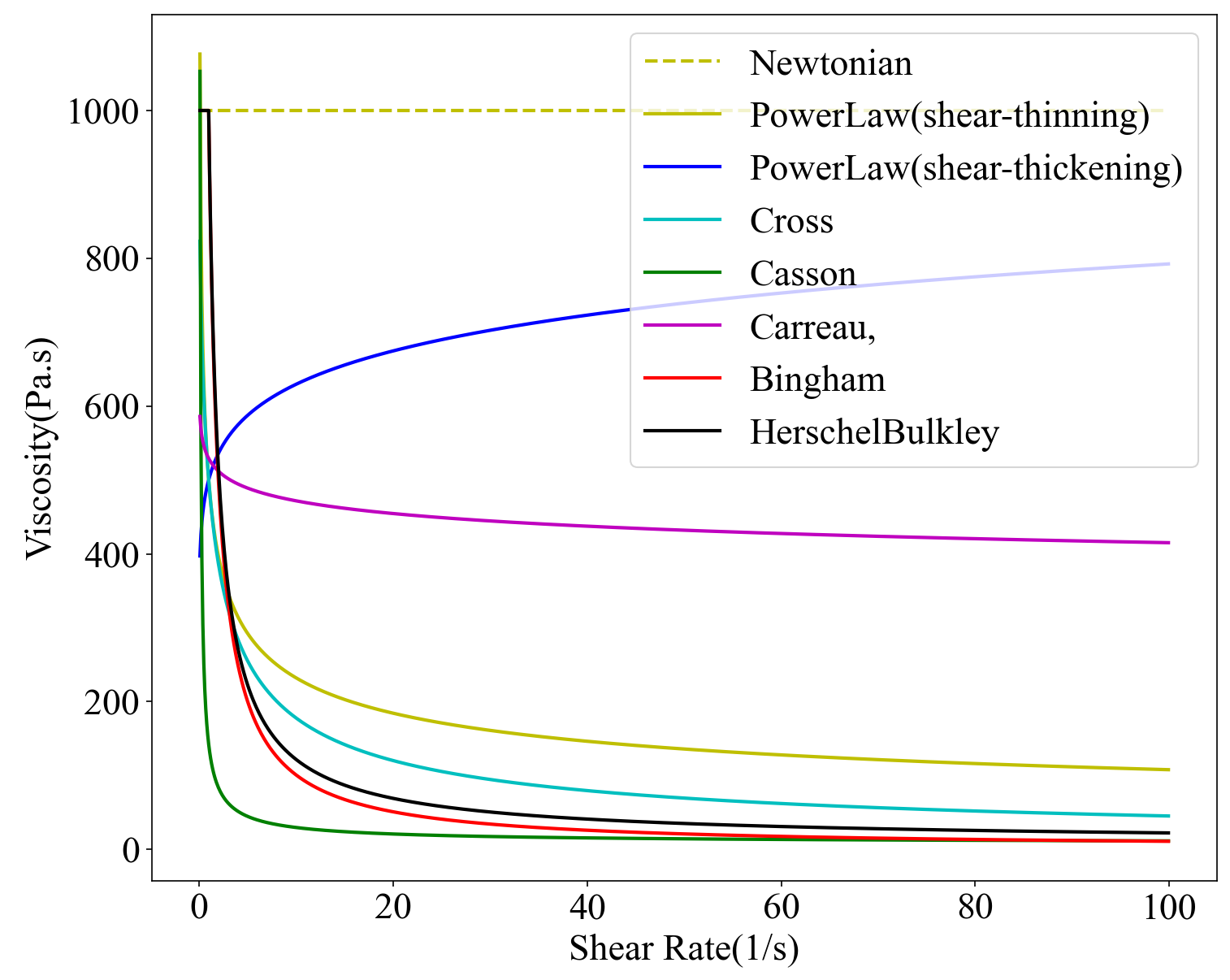}
	\caption{Different strain rate dependent viscous models. The apparent viscosity changes with the shear rate. \revised{The parameters used in Fig. 5 are listed in Table 4.}}     \label{fig:6NonNewtonsLineChart}
\end{figure}

We use the velocity gradient to calculate the strain rate, as shown in Eq.~\ref{eq:shearStrainRate}:
\begin{equation}
	\dot{\epsilon_v}=\frac{1}{2} \left(\nabla \mathbf{v}+\nabla \mathbf{v}^T\right) \label{eq:shearStrainRate},
\end{equation}
where $v$ represents velocity, and the velocity gradient $\nabla \mathbf{v}$ can be descritized as:

\begin{equation}
	\nabla \mathbf{v}_i=\frac{1}{\rho_i} \sum_j m_j\left(\mathbf{v}_j-\mathbf{v}_i\right) \nabla W_{i j}^{\top}. \label{eq:velGradient}
\end{equation}

In the following paragraphs, we will develop customized models of the six classic time-independent non-Newtonian behaviors and express them in the form of effective viscosity. Fig.~\ref{fig:6NonNewtonsLineChart} gives us an estimation of how the effective viscosity varies with the strain rate. The parameters used in Fig.~\ref{fig:6NonNewtonsLineChart} can refer to Table~\ref{tab:armadillo-params}.

\textbf{Power Law.}
The first and simplest model is the Power Law model (also known as the Ostwald De Waele model). As the name shows, it is simply a power-law relation, given by:
\begin{equation}
	\mu(\dot{\epsilon_v}) = m (\dot\epsilon_v)^{n-1} \label{eq:powerLaw},
\end{equation}
where $n$ is the power index and $m$ is the consistency index. $0<n<1$ corresponds to the shear-thinning case, and $n>1$ corresponds to the shear-thickening case. For most shear-thinning materials, $n$ ranges from 0.3 to 0.7. One advantage of the Power Law model is that it is not only capable of simulating shear-thinning, but also capable of simulating shear-thickening.

\begin{figure}[htbp]
	\centering
	\includegraphics[width=0.8\linewidth]{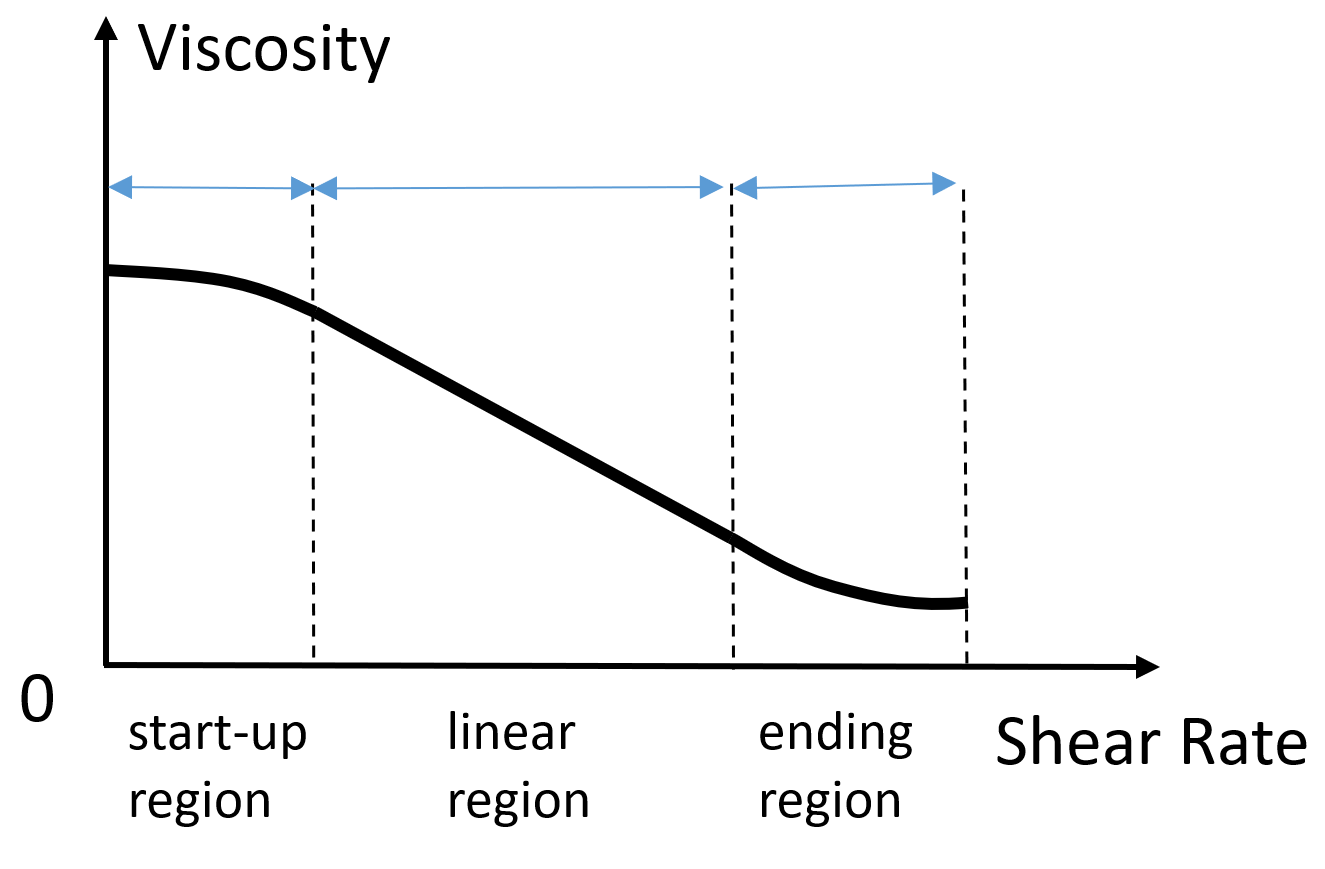}
	\caption{The reverse-S shape viscosity-shear-rate curve for Cross Model and Carreau Model}     \label{fig:reverse-S}
\end{figure}

\textbf{Cross Model.}
The Cross model is superior to Power Law for representing shear thinning behaviour with three regions, including the start-up region, the middle region, and the ending region. \revised{As shown in Fig.~\ref{fig:reverse-S}, the curve representing the relationship between viscosity and shear rate exhibits a reverse-S shape. The viscosity in the start-up region and the ending region approach $\mu_{\infty}$ and $\mu_0$, respectively. The viscosity in the middle region has a linear relation between $\mu$ and $\dot{\epsilon_v}$.} The formula is shown in Eq.~\ref{eq:cross}, where $m$ is the relaxation time. And we set $n=2/3$, which is a good default value for most materials~\cite{Cross1965}.

\begin{equation}
	\mu(\dot{\epsilon_v})=\mu_{\infty}+\frac{\mu_0-\mu_{\infty}}{1+(m \dot{\epsilon_v})^n}.
	\label{eq:cross}
\end{equation}

\textbf{Carreau Model.}
Carreau model also improves the fitting of start-up region and ending region like the Cross model~\cite{Phan2017}. As Eq.~\ref{eq:carreau} defined,

\begin{equation}
	\mu(\dot{\epsilon_v})=\mu_{\infty}+\frac{\mu_0-\mu_{\infty}}{\left(1+(m \dot{\epsilon_v})^\alpha\right)^{(1-n) / \alpha}}, \label{eq:carreau}
\end{equation}
where the default value for $\alpha$ is 2. Notably, this formula simplifies to the Newtonian model when the strain rate is extremely low and into the power law when the strain rate is extremely large.

\textbf{Bingham model.}
The Bingham model is the most straightforward model for describing a fluid of the Bingham type. These fluids are distinguished by the presence of a yield stress. When viscous stress is less than yield stress, the fluid behaves like a rigid body. When viscous stress is greater than yield stress, the Bingham fluid behaves either like Newtonian fluid or like shear-thinning fluid~\cite{Zhu2015-nonNewton}.
We adopt the Eq.~\ref{eq:Bingham} to formulate the Bingham plastic behavior without shear-thinning:

\begin{equation}
	\mu(\dot{\epsilon_v})= \begin{cases}\mu_0, & \dot{\epsilon_v} \leq \dot{\epsilon_{v,c}} \\ \mu_{\infty}+\tau_0 / \dot{\epsilon_v}, & \dot{\epsilon_v}>\dot{\epsilon_{v,c}}\end{cases}, \label{eq:Bingham}
\end{equation}
where $\mu_0$ is the initial viscosity, $\tau_0$ is the yield stress. The $\dot\epsilon_{v,c}$ is the critical strain rate corresponding to the yield stress. It must satisfy $\tau_0=\dot{\epsilon_{v,c}}\left(\mu_0-\mu_{\infty}\right)$ to make the formula continuous. A smaller $\tau_0$ means the viscosity change is more drastically.

\textbf{Casson Model.}
Casson model is a model to describe the Bingham behavior with shear-thinning. The Casson model is given by:
\begin{equation}
	\mu(\dot{\epsilon_v}) =\left(\sqrt{\mu_c}+\sqrt{ \tau_0 / \dot{\epsilon_v}}\right)^2.  \label{Casson-model}
\end{equation}
The Casson model's significant difference from the Bingham model is that there is a square root relation between effective viscosity and the strain rate.

\textbf{Herschel–Bulkley Model}
The Herschel–Bulkley model also describes Bingham's related behavior with shear thinning. It can be seen as a combination of the Power Law model and the Bingham model. The formula is given by Eq.~\ref{eq:Herschel-Bulkley}:

\begin{equation}
	\mu(\dot{\epsilon_v})= \begin{cases}\mu_0, & \dot{\epsilon_v} \leq \dot{\epsilon_{v,c}} \\ \tau_0/\dot{\epsilon_v}+m\dot{\epsilon_v}^{n-1}, & \dot{\epsilon_v} \geq \dot{\epsilon_v}_c\end{cases},
	\label{eq:Herschel-Bulkley}
\end{equation}
where it satisfies $\tau_0 = \mu_0 \dot{\epsilon_{v,c}} - m \dot{\epsilon_{v,c}}^n$ to ensure the continuity of the function.

\subsection{Elasto-plasticity}\label{sec:Elasticity-Plasticity}

In this section, we will implement the elastic and plastic models to empower our solver with the capability to solve non-Newtonian fluids to capture the time-dependent behaviors. One important characteristic of non-Newtonian material is that it neither belongs to a fluid nor a solid. It exhibits both fluid-like behavior and solid-like behavior.

Our method of simulating plasticity is improved from that of O'Brien et al.~\cite{OBrien2002} while the implementation of elasticity is based on the study of Becker et al.~\cite{Becker2009}.

\textbf{Plasticity.}\label{sec:plasticity}
When a material undergoes deformation (measured by strain) that exceeds its elastic limit, the deformation will be permanently preserved and material exhibits zero resistance to the excessive part of deformation, a phenomenon commonly referred to as yielding. The elasto-plastic strain adheres to a summation law for plastic and elastic strain, as represented in Eq.~\ref{eq:addition_law_strain}. Our primary objective is to determine the plastic strain.

First, the elastic strain deviation is calculated. It represents the strain stripped off the hydrostatic pressure, which has no impact on the plasticity.

\begin{equation}
	\mathbf{\varepsilon}^{\prime}=\mathbf{{ \varepsilon_{ep}}}-\frac{\operatorname{Tr}\left(\mathbf{{ \varepsilon_{ep}}}\right)}{3} \mathbf{I},\label{eq:elastic strain}
\end{equation}
where $\mathbf{{ \varepsilon_{ep}}}$ is the elasto-plastic strain, $I$ is the identity matrix and $\operatorname{Tr}$ is the trace.

Then we need to compare the strain deviation with the elastic limit $\gamma_1$. Only when the strain deviation exceeds $\gamma_1$, the plasticity has effect. We adopt the Von Mises criterion to judge yielding. If $\left\|\mathbf{\varepsilon}^{\prime}\right\|>\gamma_1 $ is satisfied, the yielding happens.

Additionally, the plastic strain increment is determined. Plastic strain, unlike elastic strain, accumulates from zero because plasticity is dependent on its past. Hence, we calculate the strain increment:

\begin{equation}
	\Delta \mathbf{\varepsilon}^p=\frac{\left\|\mathbf{\varepsilon}^{\prime}\right\|-\gamma_1}{\left\|\mathbf{\varepsilon}^{\prime}\right\|} \mathbf{\varepsilon}^{\prime}. \label{eq:plastic-strain-increment}
\end{equation}

Finally, we update the plastic strain:
\begin{equation}
	\mathbf{\varepsilon}^p=\left(\mathbf{\varepsilon}^p+\Delta \mathbf{\varepsilon}^p\right) \min \left(1, \frac{\gamma_2}{\left\|\mathbf{\varepsilon}^p+\Delta \mathbf{\varepsilon}^p\right\|}\right),\label{eq:plastic-strain-accumulated}
\end{equation}
where $\mathbf{\varepsilon}^p$ represents plastic strain.
The plastic limit is denoted as $\gamma_2$. Plastic strain is bounded by the plastic limit, and beyond the limit, it will tear the object into separate parts.

\textbf{Elasticity.}
Referring to Eq. \ref{eq:addition_law_strain} as we discussing the Generalized Maxwell Model, the elastic strain is obtained by subtracting the plastic strain from the elasto-plastic strain.  So this time our goal is to first calculate the elasto-plastic strain, and elastic force subsequently. We implement a linear corotated elasticity as Becker's~\cite{Becker2009}.

\begin{algorithm}
	\begin{algorithmic}[1]
		\caption{Calculate elastic force with the plasticity}\label{alg:elasticity}
		% \State \textbf{Input:} rest and current positions $x_i^0$, $x_i$
		\Statex \textbf{Input:} rest and current positions $x_i^0$, $x_i$
		\State initial neighbors and initial volumes
		\State (initial preparation)
		\For{all particles $i$}
		\State store rest position and initial neighbors
		\State set $\varepsilon_p$ to zero
		\EndFor
		\State
		\State (for every timestep)
		\For{all particles $i$} \Comment{extract rotation}
		\State compute rotation matrix $R_i$ by Eq.\ref{eq:rotation_matrix}
		\EndFor
		\For{all particles $i$} \Comment{compute stress}
		\For{all $i's$ neighbor $j$}
		\State calculate the displacement $\mathbf{u}_{ji}$ by Eq.~\ref{eq:displacement-uji}
		\EndFor
		\State calculate displacement gradient $\nabla \mathbf{u}_i$ by Eq.~\ref{eq:displacement-gradient}
		\State calculate elasto-plastic strain $\mathbf{{ \varepsilon_{ep}}}$  by Eq.~\ref{eq:total-strain-eps-u-simple}
		\State calculate strain deviation $\mathbf{\varepsilon}^{\prime}$ by Eq.~\ref{eq:elastic strain}
		\If{$\left|\varepsilon^\prime\right| > \gamma_1$}  \Comment{von Mises plasticity}
		\State calculate the plastic strain increment $\Delta \mathbf{\varepsilon}_p$ by Eq.~\ref{eq:plastic-strain-increment}
		\State update the plastic strain~$\mathbf{\varepsilon}_p$ by Eq.~\ref{eq:plastic-strain-accumulated}
		\EndIf
		\State calculate elastic strain by $\varepsilon_e = { \varepsilon_{ep}} - \varepsilon_p$
		\State transfer strain to the stress: $\mathbf{\sigma_e} = \mathbf{K}: \mathbf{\varepsilon} $
		\EndFor
		\For{all particles $i$} \Comment{compute forces}
		\For{all $i's$ neighbor $j$}
		\State calculate $\mathbf{f}_{j i}$ by Eq.~\ref{eq:elastic-force-fji}
		\EndFor
		\State calculate elastic force $\mathbf{f}_{i}^e$ by Eq.~\ref{eq:elastic-force-fi}
		\EndFor
		\Statex \textbf{Output:} elastic force with plasticity $\mathbf{f}_{i}^e$
	\end{algorithmic}
\end{algorithm}

First, we calculate the rotation matrix from the particle positions of the current configuration and initial configuration. We follow the method given by Muller et al.~\cite{Muller2005-ShapeMatching}, as shown in Eq.\ref{eq:rotation_matrix}:

\begin{equation}
	\mathbf{R}_i=\mathbf{A}_{p q_i} \mathbf{S}_i^{-1}\label{eq:rotation_matrix},
\end{equation}
where $S$ is the symmetric part of deformation matrix, $S=\sqrt{A^T_{pq} A_{pq}}$. And the $\mathbf{A}_{p q_i}$ is:

\begin{equation}
	\mathbf{A}_{p q_i}=\sum_j m_j W\left(\mathbf{x}_{i j}^0, h\right)\left(\left(\mathbf{x}_j-\mathbf{x}_i\right)\left(\mathbf{x}_j^0-\mathbf{x}_i^0\right)^T\right), \label{eq:Apq}
\end{equation}
where $\mathbf{x}_j^0$ is the initial position of $j_{th}$ particle, $x_i^0$ refers to the initial position of particle $i$, and $h$ is the kernel radius.

Secondly, it is necessary to calculate the displacement $\mathbf{u}$ and its gradient $\nabla \mathbf{u}$:
\begin{equation}
	\mathbf{u}_{j i}=\mathbf{R}_i^{-1}\left(\mathbf{x}_j-\mathbf{x}_i\right)-\left(\mathbf{x}_j^0-\mathbf{x}_i^0\right), \label{eq:displacement-uji}
\end{equation}

\begin{equation}
	\nabla \mathbf{u}_i=\sum_j \tilde{v}_j \mathbf{u}_{j i} \nabla W\left(\mathbf{x}_{i j}^0, h\right)^T,
	\label{eq:displacement-gradient}
\end{equation}
where $\tilde{v}$ denotes the initial volume of particle. Eq.~\ref{eq:displacement-uji} can be viewed as a restoring of the rigid body motion. Thus, the extracted part is the displacement.

Third, we calculate the elasto-plastic strain {$\varepsilon_{ep}$} based on the gradient of displacement $\nabla \mathbf{u}$.  Here we use the the Cauchy strain(aka. linear strain or small strain) to describe the linear elasticity:

\begin{equation}
	\mathbf{\varepsilon_e} =\frac{1}{2}\left(\mathbf{F}+\mathbf{F}^T\right)-\mathbf{I}, \label{eq:small-strain-F}
\end{equation}
where the $F$ denotes the deformation gradient, which is the Jacobian matrix of displacement. $I$ denotes the identity matrix. The $F$ can be also regarded the gradient of the deformation map, i.e., $\mathbf{F}=\nabla \mathbf{x}^0+\nabla \mathbf{u}^T=\mathbf{I}+\nabla \mathbf{u}^T$.

We can further simplify the Eq.\ref{eq:small-strain-F} by substituting $\mathbf{F}$ into it. Then we get the relation between elasto-plastic strain $\mathbf{{ \varepsilon_{ep}}}$ and  gradient of displacement $\nabla \mathbf{u}$:

\begin{equation}
	\mathbf{ \varepsilon_{ep}} =\frac{1}{2}\left(\nabla \mathbf{u}^T+\left(\nabla \textbf{u}^T\right)^T\right). \label{eq:total-strain-eps-u-simple}
\end{equation}
After getting the elasto-plastic strain, the subsequent step is to calculate the plastic strain, which has been described previously. Then, we get the elastic strain by subtracting the elastic strain from the elasto-plastic strain.

The next step is to apply Hook's Law, which can transfer the strain to the stress. The Hook's Law describes a stress-strain relation as

\begin{equation}
	{
		\mathbf{\sigma_e} = \mathbf{K} : \mathbf {\varepsilon_e},
	}
	\label{eq:Hooks-law}
\end{equation}
where $K$ is the stiffness tensor, which is a fourth order tensor.

If the material is isotropic and homogeneous, it can be expressed by only two material parameters. The two parameters have multiple equivalent choices. A common choice is the Poisson ratio $\nu$ and Young's modulus $E$. The Poisson ratio is in [0, 0.5), which measures the incompressibility of a material. The Young's modulus ranges from thousands to billions, which measures the stretch resistance.

Then, we can get the elastic force $f_{ij}$ based on the stress:

\begin{equation}
	\mathbf{f}_{j i}=-\tilde{v}_i \sigma_{e,i} \mathbf{d}_{i j} \label{eq:elastic-force-fji},
\end{equation}
where $\mathbf{f_{ji}}$ is the elastic force originating from $j$ and acting on $i$. The $d_{ij}$ is calculated by

\begin{equation}
	\mathbf{d}_{i j}=\tilde{v}_j \nabla W\left(\mathbf{x}_{i j}^0, h\right),
\end{equation}
where $\tilde{v}$ denotes the initial volume of the particle.
Finally, the elasctic force particle $i$ received is:

\begin{equation}
	\mathbf{f}_{i}^e=\sum_j \frac{-\mathbf{R}_i {\mathbf{f}}_{j i}+\mathbf{R}_j {\mathbf{f}}_{i j}}{2}.
	\label{eq:elastic-force-fi}
\end{equation}
The full algorithm to calculate elasticity and plasticity is summarized in Algorithm \ref{alg:elasticity}.

\subsection{Diffusion Model}\label{sec:diffusion}
By implementing a diffusion equation, we could further achieve the melting and solidifying phenomena. In contrast to our existing non-Newtonian viscosity models, where viscosity is controlled by strain rate, the diffusion model can be viewed as a non-Newtonian phenomenon whose viscosity is controlled by temperature~\cite{GAO2017anefficient,Su2021}. Thanks to the pure Lagrangian framework, we do not need to tackle the convection because it is naturally satisfied. All we need is to implement a diffusion equation. The governing equation that controls the evolution of a scalar field (e.g. temperature)~\cite{Cleary1999-ConductionSPH, Orthmann2012-Diffusion} is:
\begin{equation}
	\frac{\partial T_i}{\partial t}=D \sum_{j \neq i} \frac{m_j}{\rho_j \rho_i}\left(T_j-T_i\right)\left|\nabla W_{i j}\right|+R,\label{eq:diffusion}
\end{equation}
where $T$ is the temperature of the particle. $D$ is the diffusion coefficient, and it controls the diffusion speed. And $R$ is the source term, which controls the heat generation. Eq.\ref{eq:diffusion} says that the diffusion is driven by the temperature difference of the field.

% ---------------------------------------------------------------------------- %
%                                section results                               %
% ---------------------------------------------------------------------------- %

\section{Experimental Results and Comparisons}

\subsection{Comparison with Existing Works}

\begin{figure}[htbp]
	\centering
	\includegraphics[width=0.9\linewidth]{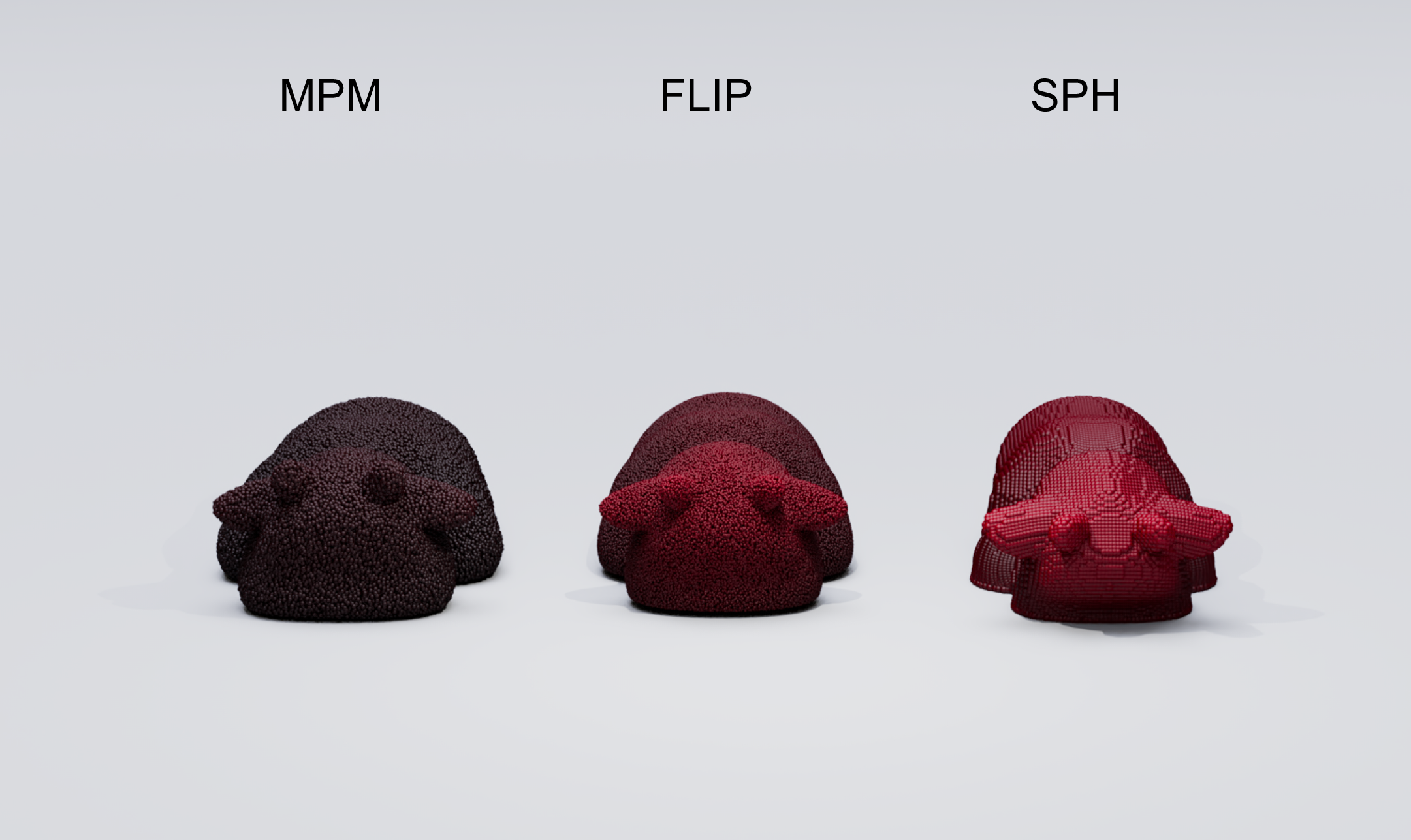}
	\caption{Comparisons of visco-elastic-plastic behaviors (particles view). From left to right, there are three toy cows modeled by MPM~\cite{Fang2019-sillyRubber}, FLIP~\cite{Shao-Huang2022-unsmoothed}, and our unified framework, respectively. The particle color represents magnitude of the instantaneous velocity.}      \label{fig:cow1}
\end{figure}

\begin{figure}[htbp]
	\centering
	\includegraphics[width=0.9\linewidth]{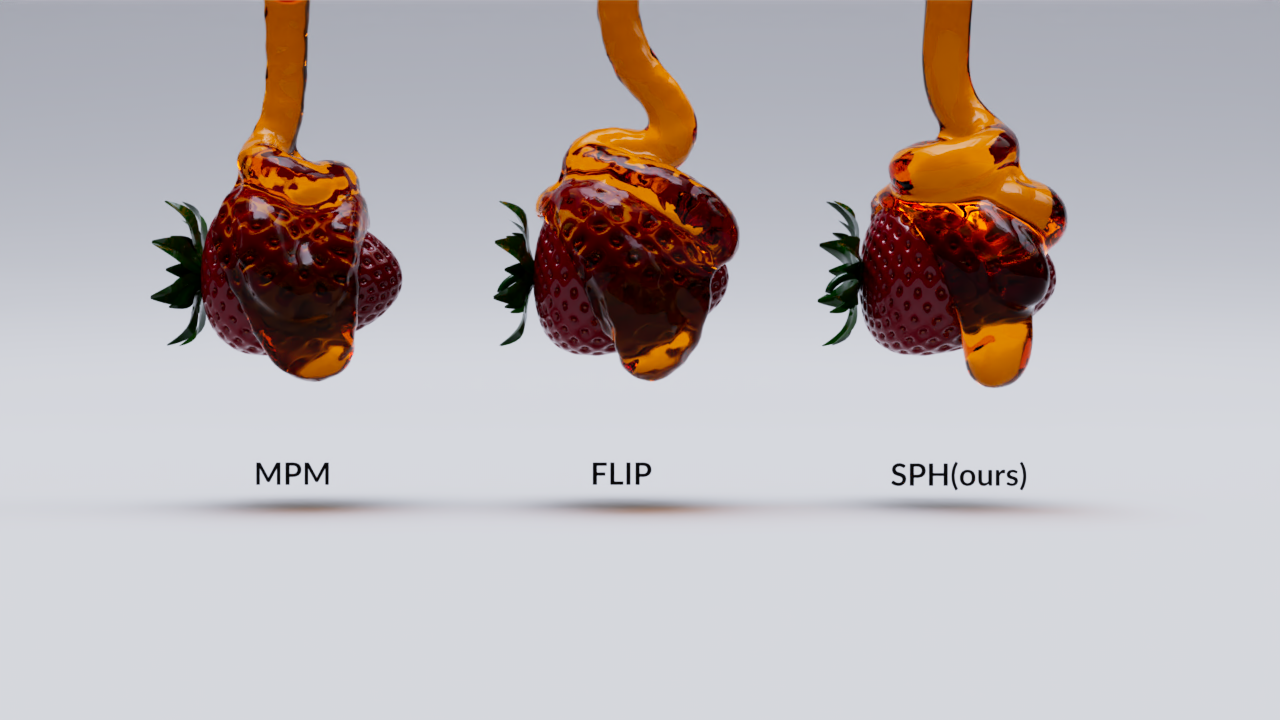}
	\caption{\revised{Comparisons of highly viscous behaviors. From left to right, there are three phenomena of honey dripping modeled by MPM (Fang et al.)~\cite{Fang2019-sillyRubber}, FLIP (Shao et al.)~\cite{Shao-Huang2022-unsmoothed} and our framework, respectively.}}     \label{fig:honey_comp1}
\end{figure}

\begin{table}[htbp]
	\centering
	\caption{\revised{Performance comparison with existing works.}} \label{tab:performance1}
	\definecolor{cyan}{rgb}{0,0,0}
	\begin{tblr}{
		width = \linewidth,
		colspec = {Q[250]Q[146]Q[204]Q[158]Q[158]},
		row{2} = {c},
		cell{1}{1} = {r=2}{},
		cell{1}{2} = {c},
		cell{1}{3} = {c},
		cell{1}{2} = {c=2}{0.35\linewidth},
		cell{1}{4} = {c=2}{0.316\linewidth},
		cell{3}{2} = {c},
		cell{3}{3} = {c},
		cell{3}{4} = {c},
		cell{3}{5} = {c},
		cell{4}{2} = {c},
		cell{4}{3} = {c},
		cell{4}{4} = {c},
		cell{4}{5} = {c},
		cell{5}{2} = {c},
		cell{5}{3} = {c},
		cell{5}{4} = {c},
		cell{5}{5} = {c},
		hline{1} = {-}{},
		hline{2} = {2-5}{cyan},
		hline{3,6} = {-}{cyan},
			}
		\textbf{Method}                                & \textbf{Time(s)/frame} &                                & \textbf{Particles number} &                                  \\
		                                               & \textbf{Cow}           & \textbf{Honey}                 & \textbf{Cow}              & \textbf{Honey}                   \\

		\textbf{FLIP}~\cite{Shao-Huang2022-unsmoothed} & \bf \underline{0.44}  & {0.06}               & {275,556}                   & \revised{13,815}                 \\
		\textbf{MPM}~\cite{Fang2019-sillyRubber}       & 23.4                   & {11.6}               & 274,462                   & \revised{13,860}                 \\
		\textbf{SPH (ours)}                            & 22.7                   & \bf \underline{\revised{0.04}} &  {276,625}   &  {\revised{13,780}}
	\end{tblr}
\end{table}

\begin{table}[htbp]
	\centering
	\caption{Supported features compared with existing works.}\label{tab:feature1}
	\begin{tabular}{llll}
		\toprule
		Framework                    & FLIP~\cite{Shao-Huang2022-unsmoothed} & MPM~\cite{Fang2019-sillyRubber} & SPH (Ours)   \\
		\midrule
		%Framework       & MPM    & MPM      & FLIP     & SPH  \\
		high viscosity               & {\Checkmark}                          & {\Checkmark}                    & {\Checkmark} \\
		\revised{variable viscosity} & {\XSolidBrush}                        & {\XSolidBrush}                  & {\Checkmark} \\
		visco-elasticiy              & {\XSolidBrush}                        & {\Checkmark}                    & {\Checkmark} \\
		plasticity                   & {\XSolidBrush}                        & {\Checkmark}                    & {\Checkmark} \\
		cutting                      & {\XSolidBrush}                        & {\XSolidBrush}                  & {\Checkmark} \\
		\bottomrule
	\end{tabular}
\end{table}

To validate the effectiveness and show the superiority of our framework, we draw comparisons with three state-of-the-art frameworks concerning highly viscous fluid and visco-elastic fluid simulation, including MPM by Fang et al.~\cite{Fang2019-sillyRubber}, FLIP by Shao et al.~\cite{Shao-Huang2022-unsmoothed}, and MPM by Su et al.~\cite{Su2021}.

Fig. \ref{fig:cow1} illustrates three toy cows with similar particles number falling on the ground, each modeled by FLIP~\cite{Shao-Huang2022-unsmoothed}, MPM~\cite{Fang2019-sillyRubber}, and our unified framework, respectively. %While FLIP is essentially a classical high-viscosity fluid simulator and does not accommodate visco-elastic-plastic modeling.
When compared with the MPM-based model of~\cite{Fang2019-sillyRubber}, which exhibits similar  visco-elastic-plastic behavior to our non-Newtonian model, our framework demonstrates superior performance.
%in terms of strain rate dependent viscosity changes. 
While the FLIP model, being a classical viscosity model, naturally exhibits the best time performance because it can not handle visco-elastic-plastic modeling. However, as shown in Table~\ref{tab:feature1}, our method can deal with more abundant non-Newtonian phenomena.

\revised{When simulating highly viscous honey, as depicted in Fig.~\ref{fig:honey_comp1}, with similar particle number, our method performs on par with FLIP~\cite{Shao-Huang2022-unsmoothed} and outperforms MPM~\cite{Fang2019-sillyRubber}.}

\begin{figure}[htbp]
	\centering
	\includegraphics[width=0.9\linewidth]{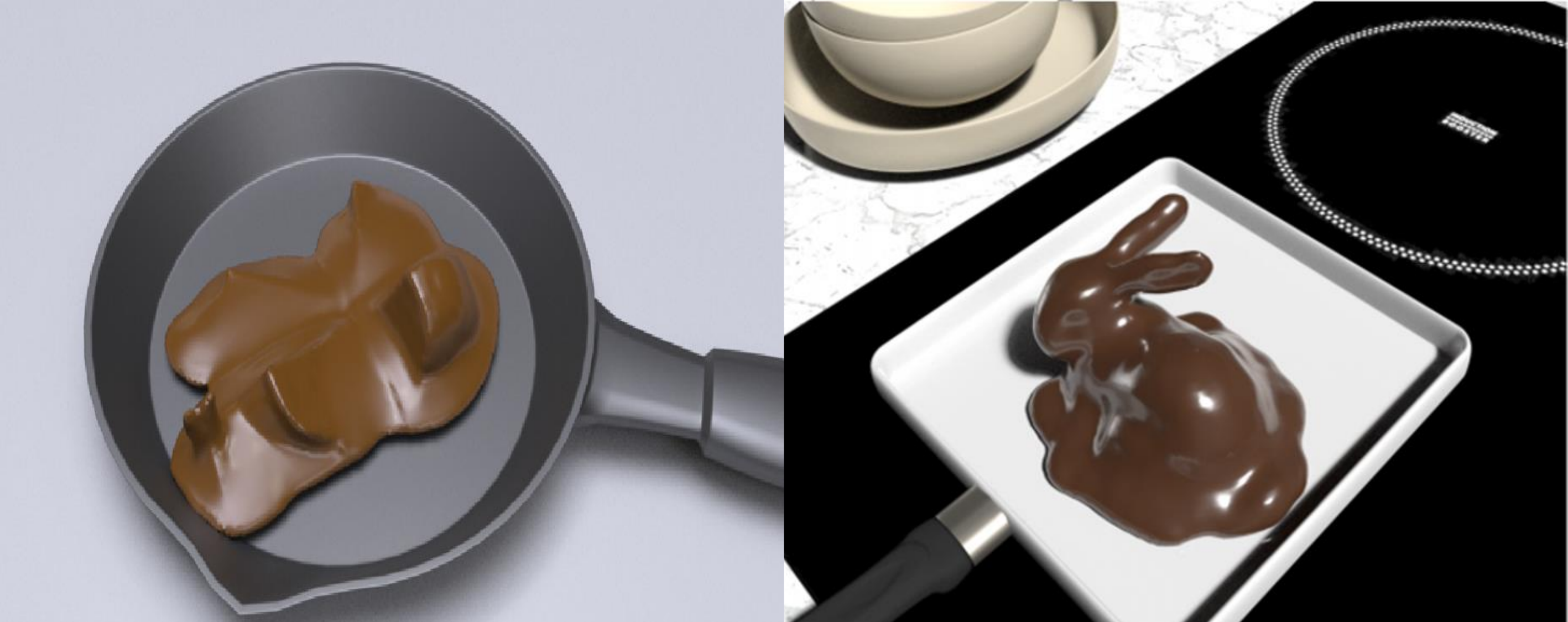}
	\caption{Comparison with ~\cite{Su2021}. Our method (left) is capable of simulating melting and cutting, with higher allowable time step size (5e-3s vs. 1e-4s). }     \label{fig:comp-melting-bunny1}
\end{figure}

Also, we have performed a comparison with Su21~\cite{Su2021} while simulating the classical melting bunny scene (as shown in Fig. \ref{fig:comp-melting-bunny1}). In contrast to this scene, our framework accommodates a larger time step size (our average time step size is 5e-3, compared to 1e-4 in~\cite{Su2021}). Moreover, owing to the pure particle framework, our work effectively handles cutting without any additional treatment.  Thanks to our unified particle system, cutting an object is a seamless process without any further complications related to topology changes.

\subsection{Phase Change of Non-Newtonian Materials}

\begin{figure}[htb]
	\centering
	\includegraphics[width=0.9\linewidth]{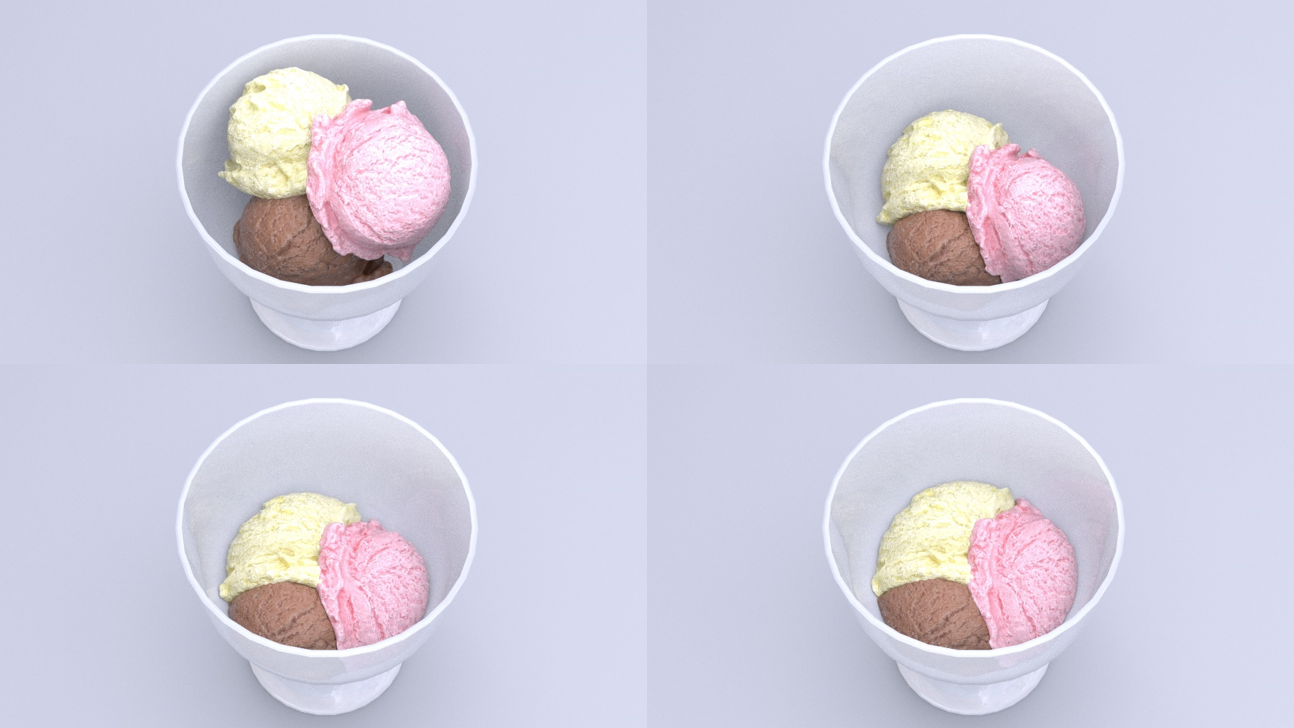}
	\caption{Ice-cream melting. }     \label{fig:ice-cream}
\end{figure}

\textbf{Ice-cream melting.}
Fig.\ref{fig:ice-cream} shows the process of non-Newtonian viscous ice-cream melting. We first set a heat source at the surface of the ice cream, and then the heat diffuses gradually into the inside, which can be controlled by Eq.\ref{eq:diffusion}. To implement this, we define the surface particles as those having ten or fewer neighbours. The viscosity of the ice cream decreases as the temperature rises. We introduce a user-controllable parameter $d$ (decay) to slow down the viscosity drop. The relation between visocisty coeffient and temperature is given by:

\begin{equation}
	\mu_i = \mu_0 e^{-d T},
\end{equation}
where $T$ is the temperature, heat source $R = 1.0$ (only at surface), $\mu_0 = 1000$ is the initial viscosity, $D = 30.0$ is the diffusivity, and decay factor $d= 0.1$.

To texturize a topologically changed animation, the texture is first mapped onto the point and carried throughout the simulation before being mapped back to the surface during rendering.

\textbf{Cutting and melting.}
The left figure in Fig.~\ref{fig:comp-melting-bunny1} illustrates the process of melting. The figure demonstrates that our solver is capable of handling object cutting as well. \revised{The cutting phenomenon is realized by the insertion of the cutting board, which is composed of particles and has a certain thickness. The board divides the bunny particles to disperse, leading them to travel beyond the neighbourhood search threshold, thereby creating the separating effect.} In this particular scene, the heat emanates from the ground (set by a bounding box with [-1,0, 1] to [1, 0.05, 1]) where a heat source $R=1.0$ is present, with a diffusivity of 100.0. And initial viscosity $\mu_0=20.0$. We use the same decaying factor $d= 0.1$.

\subsection{Shear Dependent Viscosity}
\textbf{Cornstarch Mixture (Golf Smash)}.
As shown in Fig.~\ref{fig:realworld}, we use the Power Law model in Eq.~\ref{eq:powerLaw} with a power index of 1.1 and a consistency index of 20 to simulate the scenario of a golf ball dropping into a cornstarch-water mixture. These values are chosen to ensure that the ball can naturally sink in the fluid at an initial velocity of zero.

\begin{figure}[htbp]
	%\centering
	\includegraphics[width=0.45\textwidth]{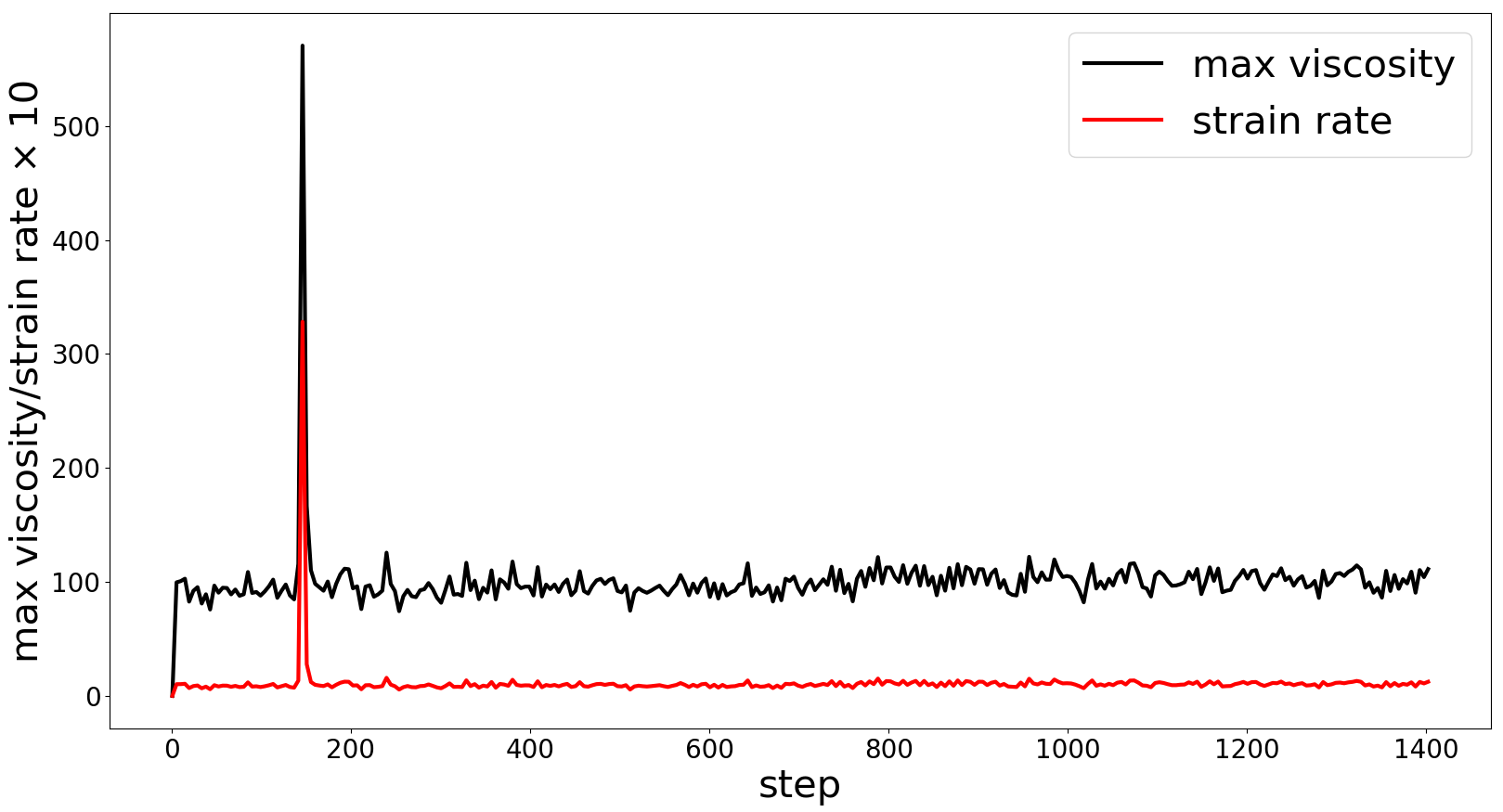}
	\caption{Max viscosity and strain rate norm in every step. Through the change of viscosity and strain rate, we can also clearly see the rigidity change at the moment of golf ball contact.
	}     \label{fig:dropline}
\end{figure}

\begin{figure*}[htbp]
	\centering
	\includegraphics[width=0.9\textwidth]{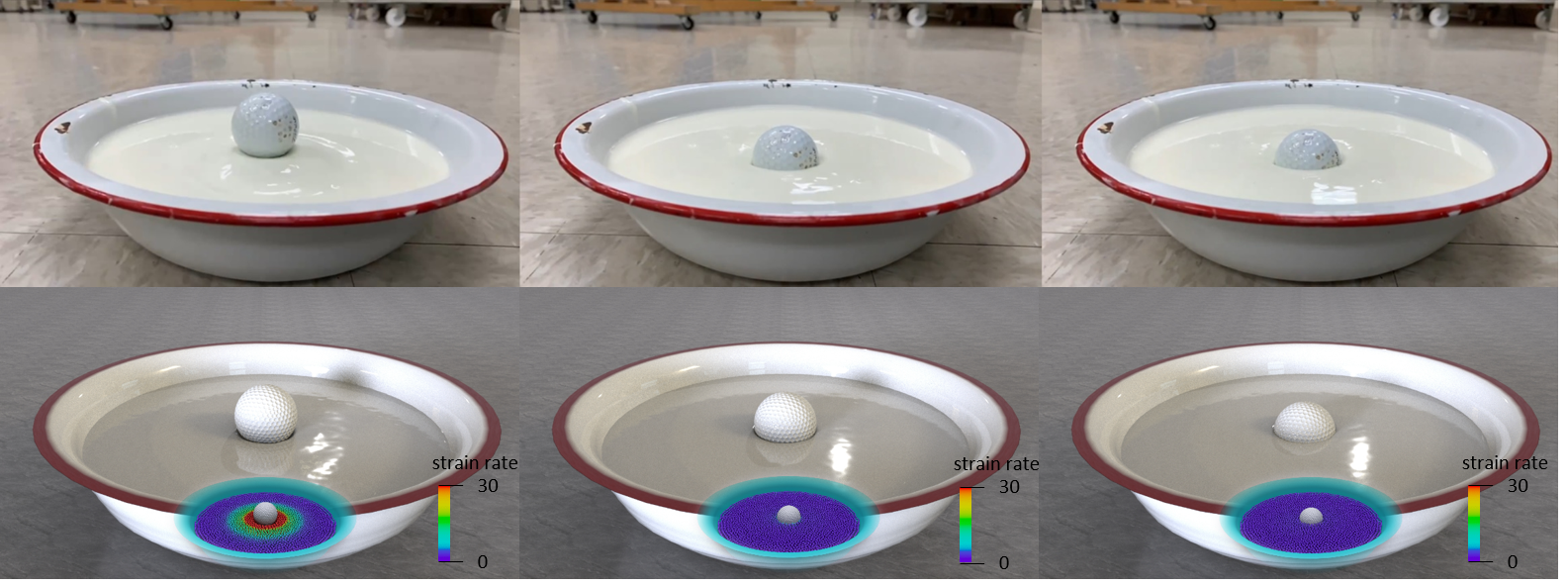}
	\caption{Real and simulated classic non-Newtonian fluid scenarios. A golf ball drops into a shear-thickening fluid of cornstarch-water mixture. The upper-row pictures are real photos and the lower ones are our simulated rendering results and particles views, where the color corresponds to the strain rate of Fig. \ref{fig:dropline}. }  \label{fig:realworld}
\end{figure*}

\begin{figure*}[htbp]
	\centering
	\includegraphics[width=.9\textwidth]{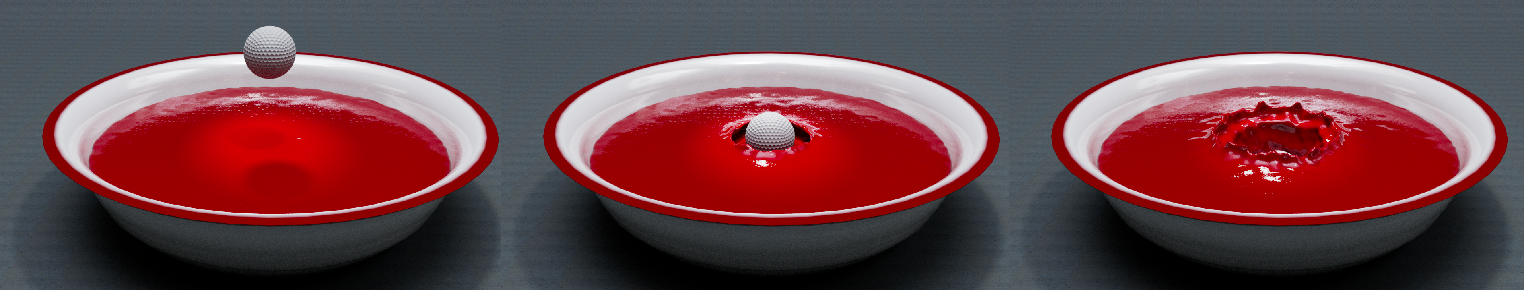}
	\caption{\revised{Change the fluid to ketchup (which is shear-thinning) and keep other setups the same with golf smashing cornstarch-water mixture case (Fig. 12).}} \label{fig:ketchup}
\end{figure*}

In this experiment(Fig.~\ref{fig:realworld}), the classic shear-thickening phenomenon is replicated. The viscosity of the fluid in  will increase with the strain rate during the simulation, resulting in a large resistance when the ball smashes into the fluid (as shown in Fig.\ref{fig:dropline}). As a result, the ball will initially slow down and then gradually drop underwater. We compare our simulation with a real-world video of a similar scenario, and the motion patterns are quite similar (Fig.~\ref{fig:realworld}).

\revised{ To compare the shear-thinning and shear-thickening behaviors, we substitute the fluid with ketchup (a shear-thinning fluid) and a golf ball falls into ketchup, while retaining other setup parameters. We adapt its parameters from~\cite{Bottiglieri1991}, using a power law model (n=0.26, m=5.86, parameters can be found in Table~3). Due to its shear-thinning nature, the viscosity decreases at the moment of collision, resulting in a splashing effect.}

\begin{table*}[htbp]
	\caption{Key parameters of our experiments.}
	\centering
	\begin{tabular}{ll}
		\toprule\label{tab:parameters}
		Scene                                                            & Parameters                                                                                \\
		\midrule
		Golf(slowly sink, Fig.~\ref{fig:realworld})                      & Power law n=1.1, m=20.                                                                    \\
		\revised{Golf(ketchup, bounce back, Fig.~\ref{fig:ketchup})} & \revised{Power law n=0.26, m=5.86~\cite{Bottiglieri1991}}                                  \\
		Armadillo (Fig.~\ref{fig:ramp})                                  & refer to Table~\ref{tab:armadillo-params}.                                                \\
		Ice Cream (Fig.~\ref{fig:ice-cream})                             & $\mu_i = \mu_0 e^{-d T}$, $\mu_0=1000, D=30, R=1, d=0.1$                                  \\
		Cut and Melt (Fig.~\ref{fig:comp-melting-bunny1})                & $\mu_i = \mu_0 e^{-d T}$, $\mu_0=20, D=1000, R=1, d=0.1$                                  \\
		Tomato(top row first, Fig.~\ref{fig:teaser})                     & $E=1e7, \nu=0.42, \gamma_1=0.001, \gamma_2 = 1.0$                                         \\
		Tomato(top row second, Fig.~\ref{fig:teaser})                    & $E=1e7, \nu=0.42, \gamma_1=0.05, \gamma_2 = 1.0$                                          \\
		Tomato(top row third, Fig.~\ref{fig:teaser})                     & $E=1e7, \nu=0.42, \gamma_1=0.1, \gamma_2 = 1.0$                                           \\
		Tomato(top row fourth, Fig.~\ref{fig:teaser})                    & $E=1e7, \nu=0.42, \gamma_1=\infty, \gamma_2 = 1.0$                                        \\
		Tomato(Newtonian, Fig.~\ref{fig:teaser})                         & $\mu_0=0.01$                                                                              \\
		Tomato(Non-Newtonian1, Fig.~\ref{fig:teaser})                    & Herschel-Bulkley $m=1.0, n=0.667, \mu_0=0.1, \dot{\varepsilon_{v,c}}=10^{-5}$             \\
		Tomato(Non-Newtonian2, Fig.~\ref{fig:teaser})                    & Herschel-Bulkley $m=20, n=1.1, \mu_0=10, \dot{\varepsilon_{v,c}}=10^{-3}$                 \\
		Tomato(Non-Newtonian3, Fig.~\ref{fig:teaser})                    & Herschel-Bulkley $m=50, n=1.1, \mu_0=100, \dot{\varepsilon_{v,c}}=100$                    \\
		Honey (our SPH, Fig.~\ref{fig:honey_comp1})                      & $\mu=5$                                                                                   \\
		Honey (FLIP, middle Fig.~\ref{fig:honey_comp1})                  & $\mu=5000$                                                                                \\
		Honey (MPM, left Fig.~\ref{fig:honey_comp1})                     & $\lambda_E=1428, \mu_E=1428, \lambda_N=28, \mu_N=7140, dx=0.05$                           \\
		Cow (our SPH, right Fig.~\ref{fig:cow1})                         & $HB, m = 1.0, n = 0.9, \mu_0=80, \dot{\varepsilon_{v,c}}=5\times 10^{-4}, \gamma_1=0.001$ \\
		Cow (FLIP, left Fig.~\ref{fig:cow1})                             & $\mu=500$                                                                                 \\
		Cow (MPM, middle Fig.~\ref{fig:cow1})                            & $\lambda_E=1428.6, \mu_E=35.7, \lambda_N=28.6, \mu_N=7.1$                                 \\
		\bottomrule
	\end{tabular}
\end{table*}

\begin{table}[]
	\caption{Parameters of the armadillo sliding scenario.}
	\centering
	\begin{tabular}{lcccccccc}
		\toprule
		                                     &
		\multicolumn{1}{l}{n}                &
		\multicolumn{1}{l}{m}                &
		\multicolumn{1}{l}{$\mu_C$}          &
		\multicolumn{1}{l}{$\tau_0$}         &
		\multicolumn{1}{l}{$\dot{\gamma_C}$} &
		\multicolumn{1}{l}{$\mu_{\infty}$}   &
		\multicolumn{1}{l}{$\mu_{0}$}                                                  \\
		\midrule
		Newtonian                            & -     & -   & -   & -  & -   & -   & 10 \\
		PowerLaw1                            & 0.667 & 4.5 & -   & -  & -   & 0.1 & 10 \\
		PowerLaw2                            & 1.5   & 1.0 & -   & -  & -   & 0.1 & 10 \\
		Cross                                & 0.667 & 1.0 & -   & -  & -   & 0.1 & 10 \\
		Casson                               & -     & -   & 1.0 & 10 & -   & 0.1 & 10 \\
		Carreau                              & 0.1   & 0.2 & -   & -  & -   & 0.1 & 10 \\
		Bingham                              & -     & -   & -   & -  & 1.0 & 0.1 & 10 \\
		HerschelBulkley                      & 0.667 & 10  & -   & -  & 10  & 0.1 & 10 \\
		\bottomrule
	\end{tabular}\label{tab:armadillo-params}
\end{table}

\begin{figure*}[htbp]
	\centering
	\includegraphics[width=0.9\textwidth]{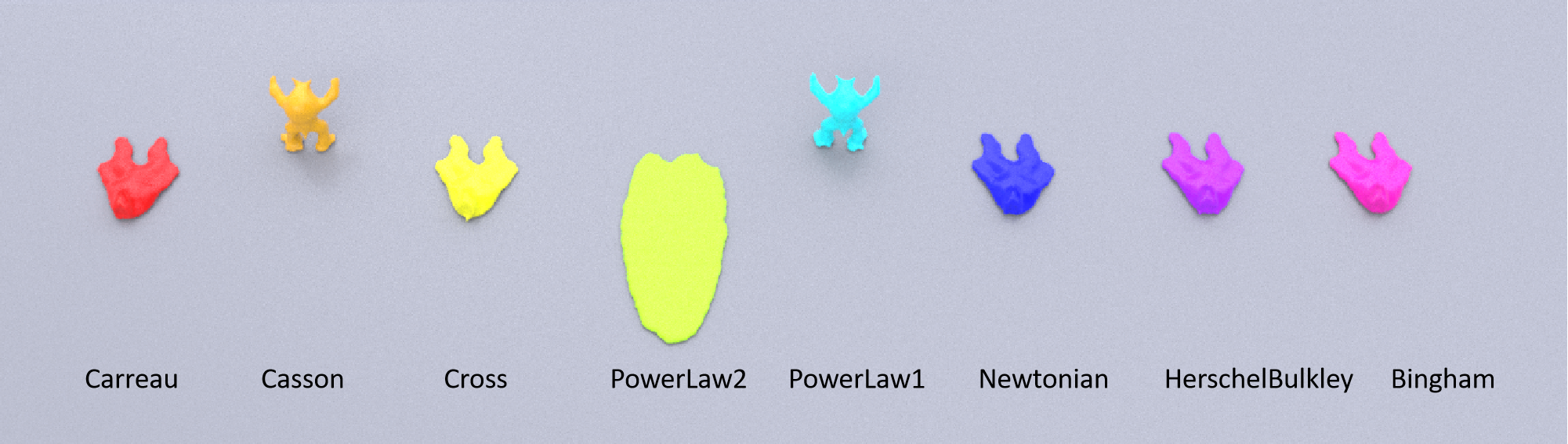}
	\caption{\revised{Armadillos with different strain-rate-dependent non-Newtonian viscosity model. From left to right: Carreau, Casson, Cross, PowerLaw2, PowerLaw1, Newtonian, Herschel-Bulkley, Bingham. Detailed parameters are listed in Table~\ref{tab:armadillo-params}.}}     \label{fig:ramp}
\end{figure*}

\textbf{Non-Newtonian Armadillos}.
Fig.\ref{fig:ramp} depicts viscous armadillos with various time-independent non-Newtonian models. Parameters are listed in Table~\ref{tab:armadillo-params}.

\revised{In this experiment, we evaluate seven distinct viscosity models that are related to shear rate. The viscosity models depicted in Fig.~14 include Carreau, Casson, Cross, PowerLaw2 (shear-thickening), PowerLaw1 (shear-thinning), Newtonian, Herschel-Bulkley, and Bingham models. These models can be classified into four categories: 1) Shear-thinning behaviors, which include Carreau, Cross, and PowerLaw1; 2) Shear-thickening behavior, primarily represented by PowerLaw2; 3) Bingham behavior, including Herschel-Bulkley, Bingham, and Casson; and 4) Newtonian behavior~\cite{Chhabra2010,Phan2017}}.

%\revised{It's important to note that the PowerLaw model can be subdivided into PowerLaw1 (shear-thinning) and PowerLaw2 (shear-thickening). This is because adjusting the parameter $n$ can emulate various non-Newtonian behaviors, such as shear-thinning ($n<1$) and shear-thickening ($n>1$). }

\revised{As shown in Fig.~14, it's clear that PowerLaw2 (shear-thickening) exhibits the optimal flowability. This is due to the fact that it simulates shear-thickening behavior, where viscosity escalates with the increase of shear rate. In our experiment, the high-viscosity fluid begins flowing from zero velocity under the influence of gravity, at a relatively low shear rate. At low shear rates, the viscosity remains low. }
\revised{In contrast, PowerLaw1 (shear-thinning) exhibits the least flowability. This is due to its simulation of shear-thinning behavior under the same parameters as PowerLaw2 (with the exception of $n$). Consequently, at low shear rate, the viscosity remains high.}

%\revised{The Bingham model demonstrates moderate flowability, similar to the Newtonian model. The defining feature of the Bingham model is that the fluid exhibits rigidity until the yield point is surpassed. Beyond that, its behavior aligns with that of a typical Newtonian fluid. For the Bingham model, the stress-strain curve beyond the yield point mirrors that of the Newtonian model.}
%\revised{It is important to note that the concept of Bingham behavior mentioned in this study is distinct from the Bingham model. The former refers to stress-strain curves with a yield point, while the latter specifically refers to the simplest model that simulates Bingham behavior, which exhibits Newtonian behavior once it exceeds the yield point.}
\revised{There are two models used for simulating Bingham behavior: the Herschel-Bulkley model and the Casson model. Beyond the yield point, the Herschel-Bulkley model adheres to PowerLaw behavior, which could manifest as shear-thinning, shear-thickening, or Newtonian. Consequently, it offers more flexibility and has a greater number of parameters. As demonstrated in Fig.~14, this model exhibits behavior similar to that of a Newtonian fluid. Conversely, the Casson model is only capable of simulating shear-thinning Bingham behavior. Therefore, in agreement with our analysis of PowerLaw1 (shear-thinning), the fluid flowability simulated by the Casson model is relatively poor at low shear rate.}

\revised{The Carreau model and the Cross model~\cite{Cross1965} are used to simulate shear-thinning behaviors. These two models are somewhat similar, with their stress-strain curves displaying a reverse S-shaped pattern. Specifically, at both low and high shear rates, the constitutive curves tend to level out. Therefore, even though both models are used to simulate shear-thinning behavior similar to PowerLaw1, the viscosity at low shear rates doesn't reach the high levels observed with PowerLaw1, resulting in moderate flowability.}

\subsection{Visco-elasticity and Visco-plasticity}\label{sec.}

\begin{figure}[htbp]
	\centering
	\includegraphics[width=.95\linewidth]{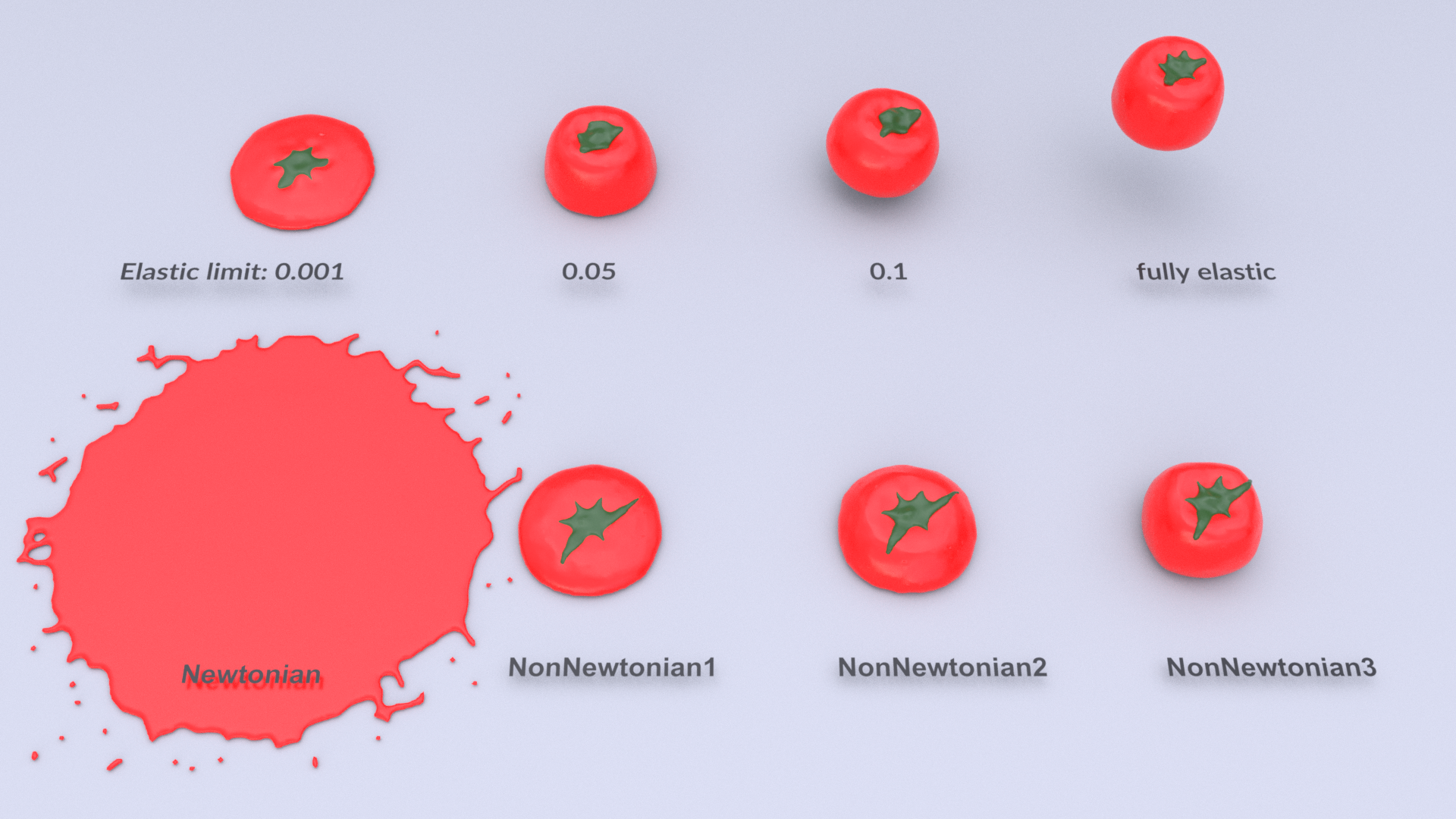}
	\caption{Various non-Newtonian tomatoes dropping on the ground. The upper row displays different levels of visco-elasticity and plasticity, ranging from more plastic to more elastic. The lower row illustrates different viscous models, progressing from less viscous to more viscous on the right. See Table \ref{tab:parameters} for parameters. }\label{fig:teaser}
\end{figure}

\textbf{Falling of tomatoes}.
Fig.~\ref{fig:teaser} shows different visco-elastic and visco-plastic tomatoes dropping on the ground.

In this scene, the Youngs modulus is 1e7, the Poisson ratio is 0.42, the plastic limit is 1.0, and the elastic limit is 0.001, 0.05, 0.1, and infinity, respectively. The upper row shows different visco-elasticity/plasticity and the bottom row shows different viscosity models.
The upper rightmost one is fully elastic and contains no plastic, therefore, the elastic limit can be seen as infinity. The existence of plasticity will make the tomato less energetic and reduce its ability to restore its original shape. As we increase the elastic limit in the upper row, which means the plastic region is easier to reach, the plastic effect will be more obvious. This gives users an easy way to control the plasticity and generate different effects as they want.

The lower row of Fig.~\ref{fig:teaser} shows non-Newtonian viscous model to exhibit that our solver is also capable of solving viscous fluid together. From left to right, the parameters  of the lower row are as follows.   Newtonian: 0.01 viscosity; Non-Newtonian1: Herschel-Bulkley model with $m=1.0, n=0.667, \mu_0=0.1, \mu_{\infty}=0.001, \dot{\gamma_C}=1e-5$, Non-Newtonian2: $m=20, n=1.1, \mu_0=10, \mu_{\infty}=0.01, \dot{\gamma_C}=1e-3$,  Non-Newtonian3: $m=50, n=1.1, \mu_0=100,\mu_{\infty}=1, \dot{\gamma_C}=100$. Readers can refer to table \ref{tab:parameters} for all parameters.

\revised{We have noticed an unnatural trembling when the elastic scene has a large number of particles (over 100K).% a phenomenon that is not observed in small-scale simulations (1K to 10K). %Adjusting the parameters has proven challenging in resolving this issue.
We suspect that this trembling is inherent to the particle-based elastic simulation itself. Given that particles are only locally connected through neighborhood searches, the information they store is confined to their immediate surroundings.} %Furthermore, compared with the mesh-based methods, particle-based methods possess more degrees of freedom, which increases with the number of particles. 
%The introduction of a damper might help mitigate this problem.}

% ---------------------------------------------------------------------------- %
%                              section Conclusion                              %
% ---------------------------------------------------------------------------- %
\section{Conclusion and Limitations}
In this paper, we develop a \revised{physically meaningful} solution for various non-Newtonian behaviors modeling. The variable physical viscous models and our empirical parameter system enable most of the non-Newtonian materials. We design a non-linear time-dependent model based on the Generalized Maxwell Model. The stress is divided into an elasto-plastic part and a viscous part, which allows us to seperatively tackle them.  The viscous and elasto-plastic stresses are combined into the Navier-Stokes equations to achieve various rheological behaviours. Meanwhile, the introduced diffusion models can support many interesting cases, such as phase change phenomenon.

There are several limitations to our method that can be addressed in future research. 
%First, although our solver is time-dependent, we do not support the time-dependent viscosity, in which resistance depends on the history of shearing. Future research can incorporate an additional parameter to describe the extent of the dynamic balance, similar to reversible chemical reactions. Readers are recommond to refer~\cite{Chhabra2010} and~\cite{Phan2017} for further details. 
First, we face the particle deficiency problem that is common to all SPH-based methods. It affects not only the quality of our simulation but also imposes numerical stability issues, especially when we need a drastic change in viscosity. We address this issue by prolonging the number of iterations of the viscosity solver, but this is an ad-hoc solution that sacrifices performance. We expect better solutions to this problem to emerge from future research. Second, we have not yet optimised the efficiency of our solver. We use the conjugate gradient as our viscosity solver and DFSPH as the algorithm to solve pressure, both of which are implemented on the CPU. No extra effort has been made to boost performance. 
\revised{Third, We also observed that unreasonable extreme parameters may lead to the non-convergence of numerical simulations. For example, rapidly changing viscosity and excessively large elastic modulus can cause issues such as numerical explosion.} 
%Some examples result in numerical explosions from the beginning, while others may experience a numerical explosion in the middle. The occurrence of a explosion in the middle is likely due to sharp variations in certain parameters. 
\revised{For instance, the moment the golf ball collides with the fluid in the falling golf ball scenario will trigger a sudden change in viscosity, which may result in a breakdown. Similarly, in the case of a falling elastic tomato, if the elastic modulus is set extremely high, the program will crash immediately. We speculate that these extreme situations may be attributed to numerical stiffness issues. Therefore, a stable viscosity and elastoplastic solver may be the key to resolving these problems. We anticipate further exploration in this area in future work.}
%\revised{Besides, we have noticed an unnatural trembling when the elastic scene has a large number of particles (over 100K), a phenomenon that is not observed in small-scale simulations (1K to 10K). Adjusting the parameters has proven challenging in resolving this issue. We suspect that this trembling is inherent to the particle-based elastic simulation itself. Given that particles are only locally connected through neighborhood searches, the information they store is confined to their immediate surroundings. Furthermore, compared to mesh-based methods, particle-based methods possess more degrees of freedom, which increases with the number of particles. }

\ifCLASSOPTIONcompsoc
  \section*{Acknowledgments}
\else
  \section*{Acknowledgment}
\fi

{This work was supported by the National Key R\&D Program of China (2023YFC3604504), National Natural Science Foundation of China (No. 62002010), CAMS Innovation Fund for Medical Sciences (CIFMS) under Grant (2019-I2M-5-016), and the Beijing Science and Technology Project (No. Z221100007722001, Z231100005923039). This work was also partially supported by USA NSF IIS-1715985 and USA NSF IIS-1812606 (awarded to Hong Qin).
}
% ---------------------------------------------------------------------------- %
%                                 End main body                                %
% ---------------------------------------------------------------------------- %

\ifCLASSOPTIONcaptionsoff
	\newpage
\fi

\bibliographystyle{IEEEtran}
\bibliography{reference}

% \bibliographystyle{elsarticle-num} 
% \bibliography{references}

\vspace{-10 mm}
\begin{IEEEbiography}[{\includegraphics[width=1in,height=1.25in,clip,keepaspectratio]{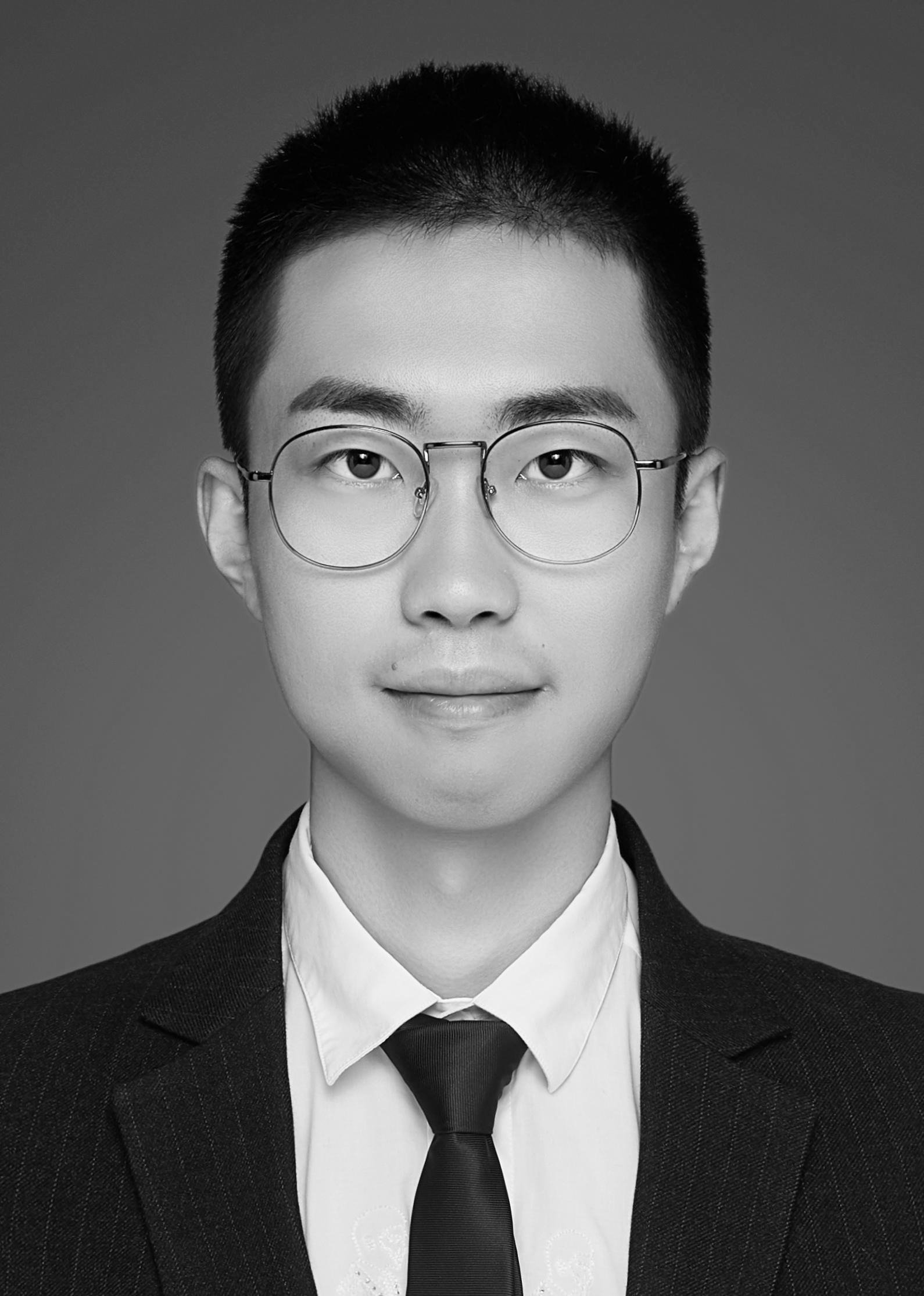}}]{Chunlei Li} received the master's degree in Power Engineering and Engineering Thermophysics from Beihang University, in 2021. He is working toward the Ph.D. degree at the State Key Laboratory of Virtual Reality Technology and Systems, Beihang University. His research interests include computer graphics and physics-based simulation. 
\end{IEEEbiography}
\vspace{-10 mm}

  \vspace{-10 mm}
\begin{IEEEbiography}[{\includegraphics[width=1in,height=1.25in,clip,keepaspectratio]{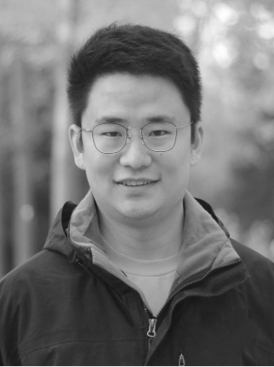}}]{Yang Gao} received the Ph.D. degree in computer science from Beihang University in 2019. He is currently an associate professor at the State Key Laboratory of Virtual Reality Technology and Systems, Beihang University, Beijing. His research interests include computer graphics, VR/AR interaction, physics-based modeling and simulation. In particular, he is focusing on fluid simulation. 
\end{IEEEbiography}   
\vspace{-10 mm}

 \vspace{-10 mm}
\begin{IEEEbiography}[{\includegraphics[width=1in,height=1.25in,clip,keepaspectratio]{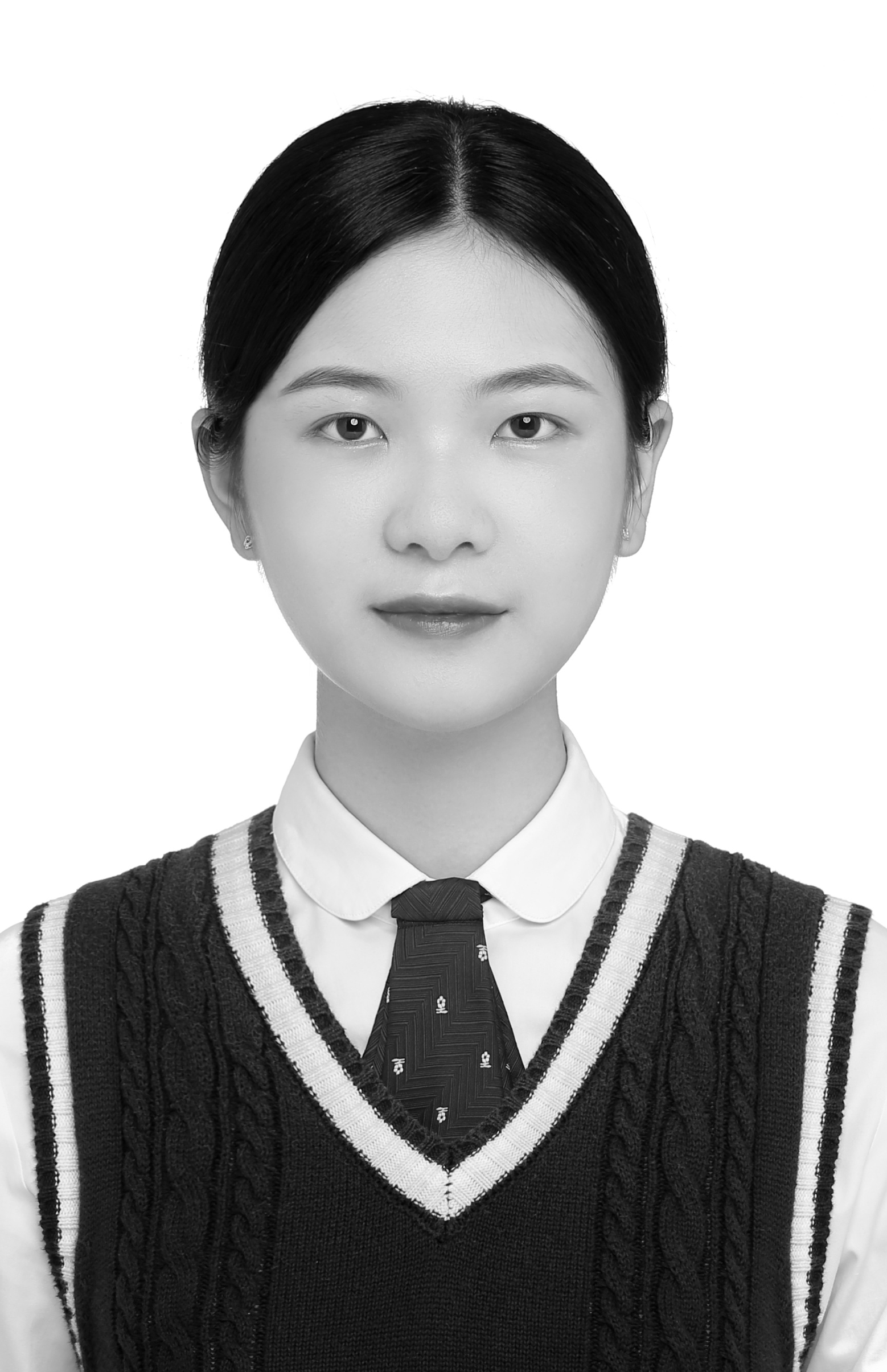}}]{Jiayi He} received her B.S. degree in computer science from Beihang University in 2023. Her research interests include computer graphics and physics-based simulation.
\end{IEEEbiography}
\vspace{-10 mm}

 \vspace{-10 mm}
\begin{IEEEbiography}[{\includegraphics[width=1in,height=1.25in,clip,keepaspectratio]{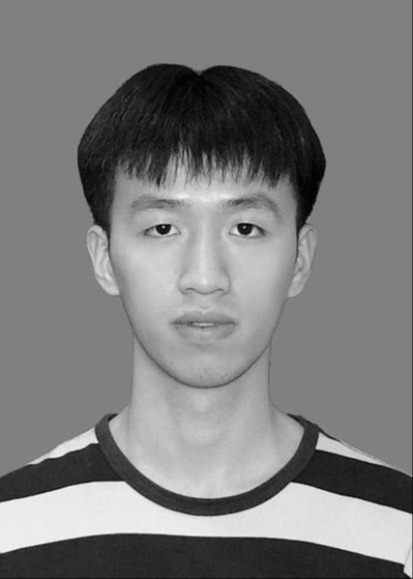}}]{Tianwei Cheng} received his master's degree in the State Key Laboratory of Virtual Reality Technology and Systems from Beihang University in 2023. His research interests include computer graphics and physics-based simulation. In particular, he is focusing on viscous fluid simulation. 
\end{IEEEbiography}
\vspace{-10 mm}

 \vspace{-10 mm}
\begin{IEEEbiography}[{\includegraphics[width=1in,height=1.25in,clip,keepaspectratio]{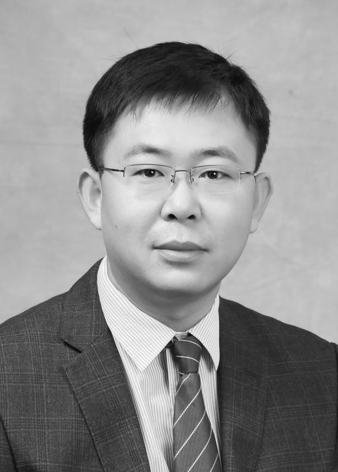}}]{Shuai Li} received the Ph.D. degree in computer science from Beihang University. He is currently a professor at the State Key Laboratory of Virtual Reality Technology and Systems, Beihang University. His research interests include computer graphics, physics-based modeling and simulation, virtual surgery simulation, computer vision, and medical image processing. 
\end{IEEEbiography}
\vspace{-10 mm}

 \vspace{-10 mm}
\begin{IEEEbiography}[{\includegraphics[width=1in,height=1.25in,clip,keepaspectratio]{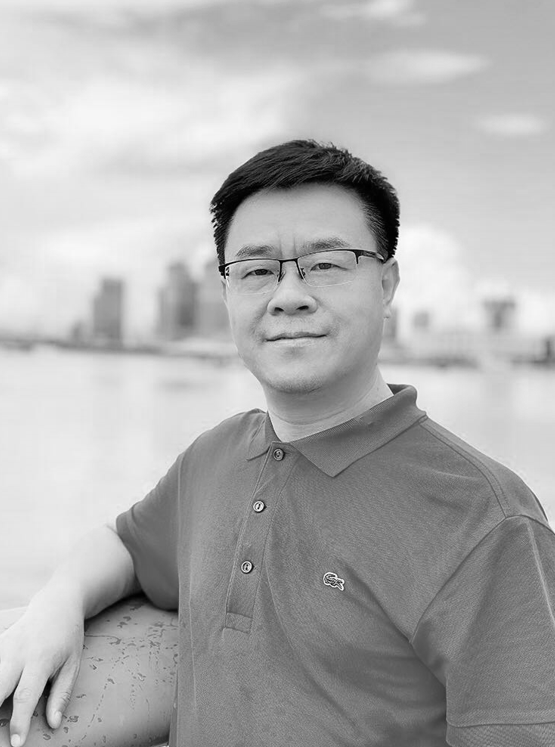}}]{Aimin Hao} is a professor in Computer Science School and the Associate Director of State Key Laboratory of Virtual Reality Technology and Systems at Beihang University. He received his B.S., M.S., and Ph.D. in Computer Science at Beihang University. His research interests are in virtual reality, computer simulation, computer graphics, geometric modeling, image processing, and computer vision. 
\end{IEEEbiography}  
\vspace{-10 mm}

 \vspace{-10 mm}
\begin{IEEEbiography}[{\includegraphics[width=1in,height=1.25in,clip,keepaspectratio]{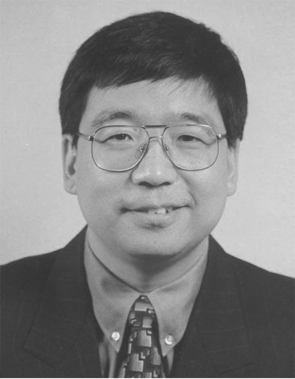}}]{Hong Qin} received his B.S. and M.S. degrees in computer science from Peking University, and his Ph.D. degree in computer science from the University of Toronto. He is a professor of computer science in Department of Computer Science at Stony Brook University. His research interests include geometric and solid modeling, graphics, physics-based modeling and simulation, computer-aided geometric design, human computer interaction, visualization, and scientific computing. He is a senior member of the IEEE and the IEEE Computer Society.
\end{IEEEbiography}

\end{document}